\documentclass[pre,twocolumn,superscriptaddress,nopacs,floatfix,amssymb,amsmath]{revtex4}
\usepackage{epsfig}
\usepackage{graphics}
\usepackage{array}
\usepackage{multirow}

\usepackage[justification=RaggedRight,singlelinecheck=false,font=small]{caption}
\usepackage{subfig}

\usepackage[usenames,dvipsnames]{color}
\usepackage{bbm}

\definecolor{MyWhite}{RGB}{255,250,240} 
\definecolor{light-blue}{RGB}{240,248,255}
\definecolor{light-gray}{gray}{0.95}

\newcommand{\beqn}{\begin{eqnarray}}
\newcommand{\eeqn}{\end{eqnarray}}

\begin{document}

\title{On phase diagram and the pseudogap state in a linear chiral homopolymer model
}

\vskip 5.0cm
\author{A. Sinelnikova}
\email{ann.sinelnikova@gmail.com}
\affiliation{Institute for Theoretical Problems of Microphysics,
Moscow State University, Moscow, 119899 Russia}
\author{A.J. Niemi}
\email{Antti.Niemi@physics.uu.se}
\homepage{http://www.folding-protein.org}
\affiliation{Department of Physics and Astronomy, Uppsala University,
P.O. Box 803, S-75108, Uppsala, Sweden}
\affiliation{
Laboratoire de Mathematiques et Physique Theorique
CNRS UMR 6083, F\'ed\'eration Denis Poisson, Universit\'e de Tours,
Parc de Grandmont, F37200, Tours, France}
\affiliation{Department of Physics, Beijing Institute of Technology,
Haidian District, Beijing 100081, P. R. China}
\author{M. Ulybyshev}
\email{ulybyshev@goa.bog.msu.ru}
\affiliation{Institute of Theoretical Physics, University of Regensburg,
D-93053 Germany, Regensburg, Universitatsstrasse 31}
\affiliation{Institute for Theoretical Problems of Microphysics,
Moscow State University, Moscow, 119899 Russia}
\affiliation{ITEP, B. Cheremushkinskaya str. 25, Moscow, 117218 Russia}

\begin{abstract}
The phase structure of a single self-interacting  homopolymer chain is investigated in terms of a universal theoretical model, 
designed to describe the chain in the infrared limit of slow spatial variations. The effects of chirality are
studied and compared with the influence of a short-range attractive interaction between monomers,
at various ambient temperature values. 
In the high temperature limit the homopolymer chain is in the self-avoiding random walk phase. At very 
low temperatures two different phases are possible: When short-range attractive interactions dominate over chirality,
the  chain collapses into a space-filling conformation. But  when the attractive interactions weaken, there
is a low temperature unfolding transition and the chain becomes like a straight rod.  
Between the high temperature and low temperature limits, several intermediate states are observed, including   $\theta$-regime and pseudogap state, which is a novel form of phase state in the context of polymer chains. Applications to polymers and proteins, in particular
collagen, are suggested. 
\end{abstract}


\maketitle

\section{Introduction}

A linear homopolymer is made of a single type of a 
repeat unit.  An important  example is polyacetylene,  an organic conductive
polymer which is the paradigm material for fractional fermion number \cite{heeger-2001,niemi-1986}.
Additional examples, among many others, are poly-L-lysine and poly-L-glutamic. 
The former is a food preservative with potential for wider, even pharmaceutically relevant
antimicrobial  effects \cite{hiraki-1995} while the latter 
%
%
is used for drug delivery against cancer \cite{li-1998}. 
%

From a theoretical point of view, the concept of a homopolymer chain is a useful 
coarse grained approximation, 
even in the case of a heteropolymer that exhibits only approximatively 
repeating patterns: For a sufficiently long  chain  the distinct monomers are simply 
combined into appropriate subunits, to dispose of  the inhomogeneities in the monomer species. 
For example collagen, which is the most 
abundant protein in mammals, displays a repeated glycine-proline-X pattern, where X is any 
amino acid other than glycine and proline. DNA, RNA and the C$\alpha$ backbone of a protein chain
are additional examples where a homopolymer approximation is occasionally profitably introduced 
\cite{degennes-1979,grosberg-1994,schafer-1999}.

Here the phase structure of  a {\it chiral} linear homopolymer  is investigated,  in terms of a
universal energy function \cite{niemi-2003,danielsson-2010,hu-2013,ioannidou-2014,niemi-2014a,niemi-2014b}; 
a chiral polymer is one where parity is broken, the mirror image 
of a stable chiral polymer conformation is in general not stable.  A chiral polymer often has a tendency 
to form helical structures. For example in the case of proteins right-handed 
helical structures are more common than left-handed ones. There are also chiral proteins 
that can form different right-handed and left-handed structures.
An important example is collagen for which the right-handed polyproline I conformation 
is more compact than the left-handed polyproline II.

It is found that  the phase structure of a chiral homopolymer is more complex than 
that of a non-chiral one. In particular, a chiral homopolymer can be in a {\it pseudogap state}. Here we give a brief description of this phase, following \cite{emery-1995,corson-1999,mannella-2005,kleinert-1998,zarembo-2002} for some abstract statistical system which displays a continuous symmetry that is associated with a complex order parameter
\begin{equation}
\Psi = \rho e^{i\varphi}
\label{Psirho}
\end{equation}
Here $\rho$ and $\varphi$ are just modulus and phase of this abstract order parameter. 
Such an order parameter is often present  {\it e.g.}  in models of superconductivity. 
Further in chapter IIE (see eq. (\ref{hmd1})) we give description of this complex order parameter in terms of polymer degrees of freedom.
The restoration of the continuous symmetry commonly  takes place so that
the free energy of the symmetry breaking state with non-vanishing
condensate 
$
\rho \ \not= \ 0
$
becomes larger than the free energy of the symmetric state where $\rho$ vanishes. But 
the symmetry can also be restored by phase decoherence, even when $\rho \not= 0$. 
This occurs when the phase $\varphi$ in the  order parameter becomes disordered so that
$
\langle e^{i\varphi}\rangle \ =  \ 0.
$
This implies that the expectation value of the order parameter also vanishes,
$
\langle\Psi\rangle \ = \ 0
$.
The system is then  
in the pseudogap state \cite{emery-1995,corson-1999,mannella-2005,kleinert-1998,zarembo-2002}.
The pseudogap state is a symmetric phase precursor 
state in the broken symmetry phase. In particular, the transition between the broken 
symmetry phase and the pseudogap state is not a phase transition but a cross-over 
prelude to the fully symmetric state, that the system enters when the lowest energy state of the effective potential
is one where the modulus vanishes.

It is reminded that the  general arguments due to  Kadanoff and Wilson 
\cite{kadanoff-1966,wilson-1971,wilson-1974,widom-1965,fisher-1975}
imply, that in the thermodynamical 
limit the phase transition properties of a material system are commonly {\it universal},  {\it i.e.} 
independent of the atomic level details. From this perspective,  the construction  of the phase 
diagram of a linear and chiral structureless homopolymer presented here, should be relevant for the 
understanding of the phase diagram of more elaborate linear chiral homopolymers, maybe 
even that of certain heteropolymers \cite{degennes-1979,grosberg-1994,schafer-1999}.
Indeed, linear polymers are presumed to have a very similar phase 
structure, quite independently of 
their chemical composition \cite{degennes-1979,grosberg-1994,schafer-1999} even though the  
phase where a particular polymer resides depends on many factors such as concentration, 
the quality of solvent,  ambient temperature and  pressure.

The article is organised as follows: The next Section describes the background and the methods.
The standard phase structure of linear 
polymers is first reviewed. The geometrical order parameter variables that are used to
model the free energy of a homopolymer are defined, followed by an outline how the ensuing
universal Hamiltonian 
emerges in the limit of low wavelength deformations. The order parameter that  detects the
presence of a pseudogap state is then defined.  The zero temperature ground state is identified in the
case of pure steric repulsion, and a universal attractive long-range interaction is introduced to model the effect
of hydrophobic forces in the homopolymer chain. A detailed analysis of various Monte Carlo algorithms
is presented, to identify one that is computationally 
most effective in the case of a single homopolymer chain. 
In the subsequent section the results are then described. The effect 
of the various parameters to the phase structure is revealed,
and in particular the pseudogap state is identified.  Finally, the phase diagram is 
constructed as a function of the various parameters. It is found that in the case of a chiral homopolymer
the phase diagram has a much richer structure  than in the case of a non-chiral homopolymer.

\section{Methods}

\subsection{Phases}

A review of the known phase structure of linear, non-chiral homopolymers is now presented, as a background and
motivation for the subsequent study.

Three different, universal phases
are commonly identified, and these phases  are  categorised by the way how the polymer structure  
fills the space \cite{degennes-1979,grosberg-1994,schafer-1999}:
Under poor solvent conditions or at low temperatures, when the attractive
interactions between the monomers dominate, a single polymer chain is presumed 
to collapse into a configuration which is space filling. On the other hand, in  a 
good solvent or at high 
temperatures, when the repulsive interactions dominate and cause the
chain to effectively swell,  its  geometric structure resembles that of a self-avoiding 
random walk (SAW). Between the two, there is a $\theta$-regime (possibly a tri-critical 
$\theta$-point) where the 
attractive and repulsive interactions cancel each other. In the $\theta$-regime 
the polymer chain is presumed to have the characteristics  of an ordinary random walk (RW). 
Finally, some polymers such as collagen for example,  are more like straight, rigid rods.
Each of these four phases -- rigid rod, SAW, RW and the space filling one --  can be characterised 
by the inverse of the Hausdorff dimension of the chain, called the scaling exponent  $\nu$  
\cite{degennes-1979,grosberg-1994,schafer-1999}.  This quantity is defined  by
the radius of gyration
\begin{equation}
R_{gyr} \ = \ \sqrt{ \, \frac{1}{2N^2}  \sum_{i,j} ( {\bf r}_i  - {\bf r}_j )^2\, }   
\label{Rg}
\end{equation}
where the $ {\bf r}_i $  are the coordinates of the individual monomers.  
When the number of monomers $N$ becomes 
very large, the radius of gyration has the asymptotic expansion
\cite{degennes-1972,leguillou-1980,li-1995,krokhotin-2012}
\begin{equation}
R^2_{gyr} \ \buildrel{N \ {\rm large}}\over{\longrightarrow}  \
R_0^2 N^{2\nu} ( 1 + R_1 N^{-\delta_1} + ... ) 
\sim \ R_0^2 N^{2\nu} 
\label{R}
\end{equation}
Here the length scale $R_0$,  the Kuhn length, 
is the effective distance  between the
monomers  in the large-$N$ limit.  The Kuhn length  
is not a universal quantity, its value can in principle
be  computed  from  the atomic level details of the polymer and 
environment including  pressure, temperature 
and chemical microstructure of the solvent. 
The dimensionless scaling exponent  $\nu$ 
{\it i.e.} the inverse Hausdorff dimension that  governs  the large-$N$ 
asymptotic form of equation (\ref{R}),  
is presumed to be a universal quantity. Its  numerical value  is 
independent of the local atomic level structure of the polymer
\cite{degennes-1979,schafer-1999,degennes-1972,leguillou-1980,li-1995}. 
The $\delta_1$ {\it etc.} are critical exponents 
and the $R_1$ {\it etc.}  are  the corresponding amplitudes, and together they constitute the 
finite-size corrections. The $\delta_1$ {\it etc.} are universal quantities \cite{li-1995}, but the $R_1$ {\it etc.}
are not universal \cite{li-1995}.

The following mean field values are conventionally assigned to $\nu$ \cite{degennes-1979,grosberg-1994,schafer-1999}: 
\begin{equation}
\nu \ =  \ \left\{  \ \, \begin{matrix} 1/3 \\ 1/2  \\ 3/5  \\ 1 \end{matrix} \right. \ \ \ \ \  \begin{matrix}
\ \,  {\rm collapsed}  \\ 
 {\rm RW}  \\  {\rm SAW}  \\  {\rm rod}  \end{matrix} 
\label{nuval}
\end{equation}
Under  poor solvent conditions or at low temperatures, the polymer 
collapses into a space-filling conformation \cite{huggins-1941,flory-1941,flory-1953} 
with the mean field exponent $\nu = 1/3$. 
Folded proteins are commonly found in this phase.
For an ordinary random-walk (RW) the mean field
value is $\nu = 1/2$. This corresponds to the  $\theta$ regime,  
that separates the collapsed phase from the high-temperature self-avoiding random walk 
phase for which the Flory value $\nu =  3/5$ is found.  Finally, when $\nu =1$, the polymer 
loses its inherently fractal  structure and behaves like a straight rod.

The transition between the collapsed phase and the SAW phase has been studied extensively, and it involves 
the RW phase as a tri-critical $\theta$-point or more  generally as a transitional $\theta$-regime. But
the transitions between the rigid rod phase and the other three, is less studied. However, there is a physically
and biologically very important scenario where such a transition could have a r\^ole; that of cold denaturation
of a protein chain. The presence of all four phases (\ref{nuval}) 
opens the possibility of a 4-critical point, 
under proper conditions \cite{bock-2007}.

Finally, the $\theta$-point value $\nu = 1/2$ is exact  for a polymer with no 
long range interactions \cite{degennes-1979}.  For a space-filling structure the value $\nu = 1/3$ is also exact, and
similarly  $\nu = 1$ is exact for a straight, linear  rod-like structure. But  in the case of SAW the 
mean field value is corrected by fluctuations.  A numerical Monte Carlo 
evaluation, computed directly by using the self-avoiding random-walk model on 
a square lattice, gives the estimate  \cite{li-1995}
\begin{equation}
\nu \  =   0.5877 \pm 0.0006.
\label{nu_SAW}
\end{equation}

In the sequel the scaling exponent  $\nu$ in (\ref{nuval}) is evaluated as a function of various parameters in different
phases of the homopolymer, using numerical simulations  in the context of a universal off-lattice  
energy function.  Of particular interest is the effect of the parameter that characterises 
the chirality, and the parameter that characterises the strength of the attractive (hydrophobic) forces.

\subsection{Geometry}

The order parameters that determine the free energy of the homopolymer chain in terms of its
geometry are now identified, following \cite{hu-2011}. For this  
a homopolymer chain with $i=1,...,N$ monomers is considered, with $ {\bf r}_i $ the three 
dimensional space coordinates.
The unit tangent vectors along the lines that connect two consecutive monomers are
\begin{equation}
\mathbf t_i = \frac{ {\bf r}_{i+1} - {\bf r}_i  }{ |  {\bf r}_{i+1} - {\bf r}_i | }
\label{t}
\end{equation}
The unit binormal vectors are defined by
\begin{equation}
\mathbf b_i = \frac{ {\mathbf t}_{i-1} \times {\mathbf t}_i  }{  |  {\mathbf t}_{i-1} \times {\mathbf t}_i  | }
\label{b}
\end{equation}
The unit normal vectors are defined by
\begin{equation}
\mathbf n_i = \mathbf b_i \times \mathbf t_i
\label{n}
\end{equation}
The three vectors ($\mathbf n_i, \mathbf b_i , \mathbf t_i$) determine an orthonormal frame at the monomer
position
$\mathbf r_i$. The discrete bond angles are
\begin{equation}
\kappa_{i} \ \equiv \ \kappa_{i+1 , i} \ = \ \arccos \left( {\bf t}_{i+1} \cdot {\bf t}_i \right)
\label{bond}
\end{equation}
and the discrete torsion angles are
\begin{equation}
\tau_{i} \ \equiv \ \tau_{i+1,i} \ = \ {\rm sgn} [(\mathbf b_{i-1}\times \mathbf b_i)\cdot\mathbf t_i] 
\times\arccos\left(  {\bf b}_{i+1} \cdot {\bf b}_i \right) 
\label{tors}
\end{equation}

Conversely, when  the angles  ($\kappa_i, \tau_i$) are known the discrete Frenet equation \cite{hu-2011}
\begin{equation}
\left( \begin{matrix} {\bf n}_{i+1} \\  {\bf b }_{i+1} \\ {\bf t}_{i+1} \end{matrix} \right)
= 
\left( \begin{matrix} \cos\kappa \cos \tau & \cos\kappa \sin\tau & -\sin\kappa \\
-\sin\tau & \cos\tau & 0 \\
\sin\kappa \cos\tau & \sin\kappa \sin\tau & \cos\kappa \end{matrix}\right)_{\hskip -0.2cm i+1 , i} \cdot
\left( \begin{matrix} {\bf n}_{i} \\  {\bf b }_{i} \\ {\bf t}_{i} \end{matrix} \right) 
\label{DFE2}
\end{equation}
determines the frames iteratively, by computing  the frame at the position of the ($i+1$)$^{th}$ monomer 
from the frame at the position of the $i^{th}$ monomer. Once all the frames have been constructed, the entire
chain is obtained as follows,
\begin{equation}
\mathbf r_k = \sum_{i=0}^{k-1} |\mathbf r_{i+1} - \mathbf r_i | \cdot \mathbf t_i
\label{dffe}
\end{equation}
With no loss of generality  one can set $\mathbf r_0 = 0$, and orient $\mathbf t_0$ to 
point into the direction of the positive $z$-axis. 

A framing is necessary for  the construction of the chain from the bond and torsion angles. 
But the equation (\ref{dffe}) does not involve the vectors $\mathbf n_i$ and $\mathbf b_i$.
Thus any linear combination of these two vectors could be chosen to define a framing, 
to construct the chain from the angles. This also determines the symmetry that
enables the identification of
the pertinent order parameter  (\ref{Psirho}):
Consider a local SO(2) transformation that rotates the  frame 
($\mathbf n_i, \mathbf b_i$) by an angle 
$\Delta_i$ leaving $\mathbf t_i$ intact,
\begin{equation}
\left( \begin{matrix}
{\bf n} \\ {\bf b} \\ {\bf t} \end{matrix} \right)_{\!i} \! \!
\rightarrow   \left(\! \begin{matrix}
\cos \Delta_i & \sin \Delta_i & 0 \\
- \sin \Delta_i & \cos \Delta_i & 0 \\ 
0 & 0 & 1  \end{matrix} \right) \! \cdot \left( \begin{matrix}
{\bf n} \\ {\bf b} \\ {\bf t} \end{matrix} \right)_{\! i}\!
\label{discso2}
\end{equation}
On the Frenet frame bond and torsion angles in (\ref{DFE2}), 
this has the following effect:
\begin{equation}
\kappa_{i}  \ T^2  \ \to \  e^{\Delta_{i} T^3} ( \kappa_{i} T^2 )\,  e^{-\Delta_{i} T^3}
\label{sok}
\end{equation}
\begin{equation}
\tau_{i}  \ \to \ \tau_{i} + \Delta_{i-1} - \Delta_{i}
\label{sot}
\end{equation}
where the $(T^a)_{bc} = \epsilon_{abc}$ are the SO(3) generators
$
[ T^a , T^b] = \epsilon_{abc} T^c
$.
The range of $\tau_i$ is $[-\pi, \pi)$ {mod}($2\pi$). The equations (\ref{sok}) and (\ref{sot}) may be 
used to extend the range of the  bond angle from  $[0,\pi)$ 
to $\kappa_{i}$  into $[-\pi, \pi)$ {mod}($2\pi$). 
The extension is compensated for by the following discrete $\mathbb Z_2$ symmetry
\begin{equation}
\begin{matrix}
\ \ \ \ \ \ \ \ \ \kappa_{k} & \to  &  - \ \kappa_{k} \ \ \ \hskip 1.0cm  {\rm for \ \ all} \ \  k \geq i \\
\ \ \ \ \ \ \ \ \ \tau_{i }  & \to &  \hskip -2.5cm \tau_{i} - \pi 
\end{matrix}
\label{dsgau}
\end{equation}
that leaves the chain intact.

In the numerical simulations presented here, all the distances between nearest neighbour 
monomers  are fixed to the uniform constant value
\begin{equation}
 |\mathbf r_{i+1} - \mathbf r_i | \ = \ \delta \ = \ 3.8 \ \ {\mathrm \AA} 
\label{dist}
\end{equation}
This equals the average 
distance between two consecutive C$\alpha$ atoms along a 
protein backbone, measured in \AA ngstr\"om's.

A polymer is subject to steric constraints, due to overlapping electron clouds and various short range Born repulsions.
 Accordingly the following forbidden volume constraint is introduced,
 \begin{equation}
 |\mathbf r_{i} - \mathbf r_k | \geq  \delta \ \equiv \ 3.8 \ \ {\mathrm \AA}  \ \ \ \ {\rm for} \ \ |i-k| \geq 2 
\label{fvol}
\end{equation}
This is in line with the minimum distance observed between any two  C$\alpha$ atoms, 
in folded protein structures.
The numerical values (\ref{dist}) and  (\ref{fvol}) can both be
independently modified, with no effect to conclusions.

In the sequel only scaled dimensionless units are used, and in particular the dimensionless unit of 
length is one \AA ngstr\"om. 

\subsection{Free energy in the infrared limit}

The bond and torsion angles constitute a complete set of geometric variables,
to describe protein C$\alpha$ backbones \cite{hinsen-2013}.
Furthermore, according to Eq. (\ref{R}) and Eq. 
(\ref{nuval}) the structure dependent phase diagram of a 
homopolymer is determined by the three dimensional chain geometry.  Thus  the
bond and torsion angles are a complete set of {\it order parameters},  in the sense of 
Kadanoff and Wilson
\cite{kadanoff-1966,wilson-1971,wilson-1974,widom-1965,fisher-1975}: 
The geometrically defined, structural  phase diagram of a homopolymer can be fully 
determined by a thermodynamical free energy which is constructed from these order 
parameters only.

%
%
A detailed derivation of the free energy used here is now presented.
For this a homopolymer chain in thermal equilibrium is considered. 
Let $F$ be the ensuing thermodynamical 
Helmholtz free energy. Thus, the minimum of $F$ describes the chain configuration, 
under thermodynamical equilibrium
conditions.  The free energy 
is the sum of the internal energy $U$ and the entropy $S$, at temperature $T$
\begin{equation}
F = U - T S
\label{Helm}
\end{equation}
It is a  function of all the  inter-atomic distances  
\begin{equation}
F = F(r_{\alpha\beta})\ ; \ \ \ r_{\alpha\beta} = |\mathbf r_\alpha - \mathbf r_\beta |
\label{intat}
\end{equation}
where the indices $\alpha,\beta, ... $ extend over all the atoms in the homopolymer system, including those  
of the solvent environment.
Consider the infrared, long distance limit where the characteristic length scales of spatial 
deformations along  the homopolymer chain around its thermal equilibrium configuration are
large in comparison to the distance (\ref{dist}) between neighboring monomers. This is synonymous
to an assumption that there are no abrupt wrenches and buckles along the chain, that there are only gradual
long wavelength  bends which is the limit of adiabatic deformations. The  
completeness of the bond and torsion angles to describe a protein structure \cite{hinsen-2013}
implies that, in order to determine the thermodynamical
phase state of the homopolymer chain, it is sufficient to consider the response of all the distances 
between all the atoms to the variations in the bond and torsion angles only,
\begin{equation}
r_{\alpha\beta} \ = \  r_{\alpha\beta} (\kappa, \tau).
\label{r_kappa_tau}
\end{equation}
Here, and in the sequel, ($\kappa, \tau$) denotes collectively all the variables $\kappa_i$ and $\tau_i$.
 
Suppose that at a local extremum of the free energy, the bond and torsion angles along the homopolymer chain
have the values
\begin{equation}
(\kappa_i , \tau_i) \ = \ (\kappa_{i0},  \tau_{i0})
\label{kappa_tau0}
\end{equation}
Consider a conformation where the ($\kappa_i, \tau_i$) deviate from  these
extremum values. The deviations are
\begin{equation}
\begin{matrix} 
\Delta \kappa_i & = & \kappa_i - \kappa_{i0} \\
\Delta \tau_i & = & \tau_i - \tau_{i0}
\end{matrix}
\end{equation}
Start by Taylor expanding the infrared limit Helmholtz free energy  (\ref{Helm})  around the extremum, 
\[
F\left[ r_{\alpha\beta} =   r_{\alpha\beta} (\kappa_i, \tau_i)\right] \ \equiv \
F(\kappa, \tau) \ = \ F(\kappa_{0}, \tau_{0}) 
\]
\[
+ \sum\limits_k \left\{ \, 
\frac{\partial F}{\partial \kappa_k}_{|0}\! \Delta \kappa_k 
+ \frac{\partial F}{\partial \tau_k}_{|0}\! \Delta\tau_k \, \right\} 
\]
\[
+  \sum\limits_{k,l} \left \{ \, \frac{1}{2}
\frac{\partial^2 F}{\partial \kappa_k \partial \kappa_l }_{|0}\! \Delta \kappa_k  \Delta \kappa_l
+ \frac{\partial^2 F}{\partial \kappa_k \tau_l}_{|0}\! \Delta\kappa_k \Delta \tau_l  + \right.
\]
\begin{equation}
\left. + \frac{1}{2}
\frac{\partial^2 F}{\partial \tau_k \partial \tau_l }_{|0}\! \Delta \tau_k  \Delta \tau_l
\, \right\} + \mathcal O (\Delta^3).
\label{free_energy}
\end{equation}
The first term in the expansion evaluates the free energy at the extremum. Since ($\kappa_{i0}, \tau_{i0}$)
correspond to the extremum, the second term vanishes.  
Denote in the sequel ($\kappa_i,\tau_i$) collectively, as the variable  $\rho_i$. Then,
\[ 
F(\kappa, \tau) \ \equiv \ F(\rho)= 
\]
\begin{equation}
= \ F(\rho_{0}) +  \frac{1}{2} \sum\limits_{k,l} 
\frac{\partial^2 F}{\partial \rho_k \partial \rho_l }_{|0}\! \Delta \rho_k  \Delta \rho_l \ + \ \mathcal O (\Delta^3)
\label{Fene}
\end{equation}
Following \cite{coleman-1973} the expansion (\ref{Fene}) is {\it re-arranged}  
in terms of of the differences in the angles $\rho_i \sim$
($\kappa_i,\tau_i$), 
as follows:
\[
F(\rho) =  \sum\limits_k \left\{ \ V_k(\rho_k; \rho_{0k}) \ + \right.
\]
\begin{equation}
\left. + \  Z_k (\rho_k; \rho_{0k}) 
( \rho_k \rho_{k+1} +  \rho_k \rho_{k-1} ) \ +  \ \dots \ \right \}
\label{cwexp}
\end{equation}
Here  $\rho_{0k}$ denotes a combination of the various parameters 
($\kappa_{i0}, \tau_{i0}$) along the chain.  But $V_k(\rho_k; \rho_{0k})$, $ Z_k (\rho_k; \rho_{0k}) $
and so forth depend on the variable $\rho_k$ only on the site $k$; these functions are {\it ultralocal}.
The terms that are not shown explicitly, consist of higher order differences $\rho_k \rho_{k+i}$ with
$i \geq 2$, and higher powers of the differences. 
The local terms $V_k(\rho_k)$ constitute the {\it effective potential}
\begin{equation}
V_{eff} = \sum\limits_k  V_k(\rho_k) 
\label{Veff}
\end{equation}
The structure  of the effective potential is commonly used to conclude whether a spontaneous 
symmetry breaking takes place \cite{coleman-1973}.

The transition from (\ref{Fene}) to (\ref{cwexp}) 
involves, {\it a priori}, an {\it infinite} re-arrangement of the terms in the Taylor expansion (\ref{Fene}).
In particular, the expansion (\ref{cwexp}) has been designed  so that in the continuum limit where 
distance between neighboring monomers vanishes {\it i.e.} $\delta \to 0$ in (\ref{dist}), it becomes,
at least {\it naively}, 
an expansion of the free energy in powers of  momentum about the point 
where momentum vanishes: For a single scalar variable $\rho_k$ with continuum limit
$
\rho_k \to \phi(x)
$
the corresponding continuum limit of (\ref{cwexp}) is the derivative expansion \cite{coleman-1973} 
\begin{equation}
F(\phi) = \int \left[ V(\phi) + \frac{1}{2} (\partial_\mu \phi)^2 Z(\phi) + \dots \right]
\label{Fphi}
\end{equation}

\subsection{Effective Hamiltonian}

Clearly, the free energy must remain invariant under the local frame rotations (\ref{sok}),
(\ref{sot}); the physical properties of the chain do  not depend on the choice of framing.  
Accordingly, it has been concluded 
 \cite{niemi-2003,danielsson-2010,hu-2013,ioannidou-2014,niemi-2014a,niemi-2014b}
that -- in the unitary gauge -- 
to the leading non-trivial order the free energy has the form  
\[
H=-\sum_{i=1}^{N-1} 2\kappa_{i+1} \kappa_i  
+ \sum _{i=1}^{i=N}  \left\{ 2
\kappa_i ^2 + q  ( \kappa_i^2-m^2)^2 + cd \kappa_i^2 \tau_i^2   \right\}
\]
\begin{equation}
+  \sum _{i=1}^{i=N}  \left\{ \frac{1}{2} c \tau_i^2 - a \tau_i  - b \kappa_i^2 \tau_i \right\}
\label{hamiltonian}
\end{equation}
This  is adopted as the (effective) Hamiltonian, in the sequel.
In  (\ref{hamiltonian}) $q$, $m$,  $a$, $b$, $c$, $d$ 
depend on the atomic level physical properties and the chemical 
microstructure of the homopolymer chain and its environment. In principle, these parameters can 
be computed from this knowledge. 
(Note the combination $cd$ in the last term of second sum; this choice 
is made for later convenience.)

It can be shown
\cite{niemi-2003,danielsson-2010,hu-2013,ioannidou-2014,niemi-2014a,niemi-2014b} that
(\ref{hamiltonian}) is the most general, universal and 
gauge  {\it i.e.} frame rotation (\ref{discso2}) 
invariant Hamiltonian, that models a
homopolymer in the limit  where the characteristic length 
scales of spatial deformations around the minimum energy  
configuration become large in comparison to the distance (\ref{dist}) 
between consecutive monomers.  The effective Hamiltonian
(\ref{hamiltonian}) coincides with the {\it naively} discretized 
continuum {\it Abelian Higgs Model Hamiltonian}
with one complex scalar field, when expressed 
in the unitary gauge and  the  U(1) gauge transformation is identified
with the frame rotation (\ref{discso2}); the term with parameter $a$ is the Chern-Simons term,
commonly introduced in gauge theories to break parity.
In particular, (\ref{hamiltonian}) is {\it unique} in the sense of Kadanoff and Wilson. 

Implicit in  (\ref{hamiltonian}) is the assumption
that there are no abrupt wrenches and buckles along the polymer chain.
Only small, gradual bends are present in the deviations around the energy 
minimum configuration, which is obtained by minimising the Hamiltonian  (\ref{hamiltonian}).
This defines the limit of adiabatic deformations.

In line with the  Abelian Higgs Model, see  
\cite{chernodub-2008} in the present context, the Hamiltonian (\ref{hamiltonian}) 
displays the discrete symmetry
$
\kappa_i \to - \kappa_i
$.
As in the case of the Abelian Higgs Model, 
this symmetry may become spontaneously broken by the ground state.
It should be noted  that  (\ref{hamiltonian}) is {\it not} invariant under the 
local $\mathbb Z_2$ gauge symmetry (\ref{dsgau}), as it 
coincides with the leading  non-trivial contribution
to an expansion of the Helmholtz free energy around a fixed background. To 
recuperate the $\mathbb Z_2$ symmetry one may  replace $\tau$ in Hamiltonian by $\frac{1}{2} \sin 2\tau$.
Alternatively, the Hamiltonian (\ref{hamiltonian}) can be interpreted as 
a deformation of the standard energy function of the discrete 
nonlinear Schr\"odinger  equation (DNLS)  \cite{faddeev-1987,ablowitz-2003}. The first two sums 
coincide  with the energy of  the standard  DNLS equation,  
in terms of the discrete Hasimoto variable of \cite{hu-2013}. 
The first  ($c$) term in the third sum is the Proca mass that has a claim of gauge invariance; here the Proca
mass is a ``regulator", as explained in \cite{niemi-2014b}.
The second ($a$) term is the 
 helicity, and the last ($b$) term is the conserved momentum. The last 
 two terms break the $\mathbb Z_2$ parity 
 symmetry, these two terms are responsible for helicity of the homopolymer chain.

\vskip 0.2cm
The simulations that are described in this article have been performed by keeping some of the
parameter values fixed. In Table \ref{table-1} the parameter values that are kept fixed, 
have been listed.
\begin{table}[tbh]
\caption{The parameters in (\ref{hamiltonian}) that are kept fixed during our simulations.}
\vspace{10mm}
\begin{tabular}{|c|c|c|c|c|}
\hline 
~~~$q$~~~ & ~~~$m$~~~ & ~ b ~ & ~ c ~ & ~$d$~  \\
\hline
 $3.5$ & $1.5$ & $0$ & $10^{-4}$ & $10^{-4}$  \\
\hline
\end{tabular}
\label{table-1}
\end{table}
The numerical values of these parameters have been chosen in conformity with those, that are commonly 
encountered in the case of proteins \cite{krokhotin-2012,molkenthin-2011}; for 
a protein, the torsion angles are much more 
flexible than bond angles.  The value of $m$ in Table \ref{table-1} 
corresponds to an $\alpha$-helical 
structure. 

The parameter $a$, which is not fixed, is of particular interest in the sequel. This
is the parameter that breaks chirality; note that the momentum of the DNLS hierarchy 
is not considered 
here {\it i.e.} $b=0$ for simplicity. This term lacks a direct interpretation in the context
of the Abelian Higgs Model.  It turns out that the effects of this term are largely
accounted for by the $a$ dependent helicity, in any case.

\subsection{Pseudogap}
\label{sec:pseudogap}

In \cite{hu-2013} the  following combination of the  bond and torsion angles has been considered
\begin{equation}
\psi_i  \ = \ \sigma_ie^{i \vartheta_i} \ \equiv \ \tan \frac{\kappa_i}{2} \, e^{i \vartheta_i}
\label{hmd1}
\end{equation}
where  the phase is
\begin{equation}
\vartheta_i = \frac{1}{2} \left( \sum\limits_{k=1}^i \tau_k - \sum_{k=i+1}^N \tau_k \right).
\label{phases}
\end{equation}
The variable (\ref{hmd1}) is essentially the discrete version of the Hasimoto variable \cite{hu-2013}, 
in terms of the  Frenet 
frame coordinates: It is the complex variable that  relates  
(\ref{hamiltonian})  into a generalised version of the discrete non-linear Schr\"odinger 
equation  \cite{faddeev-1987,ablowitz-2003}. It is also the present version of the complex order parameter
(\ref{Psirho}). Thus the pseudogap state can be identified with a state where 
the bond angles are non-vanishing and ordered
\begin{equation}
\langle\kappa_i \rangle \ = \ \kappa_0 \ \not= 0
\label{broken}
\end{equation}
for some site independent $\kappa_0$, while torsion angles are essentially randomly 
fluctuating to the effect that
\begin{equation}
\langle e^{i\vartheta_i} \rangle \ \approx  \ 0.
\label{zero_order}
\end{equation}
Accordingly  in the sequel the pseudogap state is detected by monitoring both $\kappa_i$ 
and $\tau_i$ simultaneously. It should be noted that
since the effective potential (\ref{Veff})  is  insensitive
to the phase in (\ref{hmd1}), the pseudogap state
can be difficult to detect in terms of the minima of the effective potential alone. 
A dynamical computation that engages fluctuations is needed, to detect the 
presence of the pseudogap.  

It should be kept in mind, that a relation such as (\ref{broken}) is commonly deduced 
by inspection of the effective potential. In the full theory there are always corrections,
due to fluctuations. In particular, in the full theory, at a finite temperature,  (\ref{broken}) 
never  vanishes identically. The modulus of the order parameter (\ref{hmd1}) 
is a positive definite quantity and thus, due to fluctuations, it 
always acquires a non-vanishing value in the full theory, as also shown in the simulations presented here.

\subsection{Zero temperature}

The zero temperature ground state of the Hamiltonian  (\ref{hamiltonian}) is a solution to the 
equations of motion,
\begin{equation}
\tau_i \ = \ \frac{a}{c} \frac{ 1 } {d\kappa_i^2 + 1}
\label{taueq}
\end{equation}
\begin{equation}
\kappa_{i+1} \ = \ 2 \kappa_i - \kappa_{i-1} + 2 q (\kappa_i^2 - m^2) \kappa_i + c d \tau_i^2 \kappa_i
\label{kappaeq}
\end{equation}
Accordingly the minimum energy 
ground state of  (\ref{hamiltonian}) is
\begin{equation}
\begin{matrix}
\kappa_i & = & \pm\sqrt{ m^2 - \frac{cd}{2q} \tau_i^2 \, }  \ \approx \ \pm m 
\\ & & \\
\tau_i & = & \frac{a}{2c}   \frac{1}{dm^2+1}  \ \approx \  \frac{a}{2c}  
\end{matrix}
\label{ksolu}
\end{equation}
In the sequel the parameter $m$ has the fixed value, given in Table \ref{table-1},  throughout. 
As a consequence the ground state is controlled by the ratio $a/c$, and  in the sequel the 
phase structure is investigated by varying the ratio $a/c$  
within the range $a/c \in [0,4\pi]$. 

It should be noted that the configuration (\ref{ksolu}) does not necessarily describe the minimum energy
homopolymer: For some parameter values there can be a conflict between the values of ($\kappa_i,\tau_i$) 
given by  (\ref{ksolu}) and the forbidden volume constraint (\ref{fvol}). A
configuration (\ref{ksolu}) which satisfies the forbidden volume constraint is a helix. But if the forbidden volume 
constraint 
is not obeyed, the lowest energy ground state configuration is the one that minimises (\ref{hamiltonian}), 
subject to  the constraint (\ref{fvol}). 

The range of the  bond angle become extended to negative values, by the $\mathbb Z_2$ symmetry
(\ref{dsgau}).  The two ground states $\kappa_i = \pm m$ have the same energy.
In addition of these two ground states, there can also be local 
minima of (\ref{hamiltonian}) that have the profile of a kink \cite{chernodub-2010,molkenthin-2011} {\it i.e.}  
a domain wall that interpolates between the two ground states $\kappa_i = \pm m$.  
The energy of a kink is higher than the energy of  the ground state helix (\ref{ksolu}).  Two kinks can 
annihilate each other,  thus any pair of kinks can be removed by continuous deformations of the chain.  
A single kink can be translated, so that it becomes removed through the ends of the chain.  
However,   on a discrete lattice the translation invariance is commonly
broken, by the 
Peierls-Nabarro barrier \cite{peierls-1940,nabarro-1947,nabarro-1997,sieradzan-2014}.
Thus, in general it costs (thermal)
energy to translate a kink along the chain.

The present scenario is different from the one that appears in the case of kinks in folded proteins 
\cite{chernodub-2010,molkenthin-2011}.
There, the parameter values in (\ref{hamiltonian}) are different  for different 
super-secondary structures {\it i.e.} helix-loop-helix motifs. A folded protein is described by
a heteropolymer generalisation of (\ref{hamiltonian}), and  the ground state is not a straight helix
such as (\ref{ksolu}).  A short  analysis of a simple 
heteropolymer is presented in the sequel, in sub-Section \ref{sec:hetero}.



\subsection{Attractive interaction}
\label{sect:attractive}

The constraint (\ref{fvol}) is a purely repulsive interaction, and in the case of a homopolymer
it models forbidden volume constraints which have a short spatial range. It could be 
generalized to include an attractive component, with a short range but one that exceeds the extent of 
forbidden volume constraints. 
Accordingly, the following rudimental  extension of  (\ref{fvol}) to model both short range forbidden volume 
constraints and attractions is introduced. 
\begin{equation}
 U(r) \ = \left\{ { { \ + \infty \ \ \  \ \ \ \ \ \ \ \ \ \ \ \ \ \ \ \   \quad   0 < r < \delta } \atop
 {\ U_0 \! \left\{ \tanh(r-R_0)-1\right\}  \quad  \delta < r < +\infty } } \right.
\label{U}
\end{equation}
Here $\delta$ is the radius of the self-avoiding condition (\ref{fvol}).
For $r<\delta$ the forbidden volume condition (\ref{fvol}) persists. But for $r>\delta $
 there is a short range attractive interaction with 
strength determined by the parameter $U_0$; when the parameter $U_0$ 
vanishes (\ref{U}) reduces to (\ref{fvol}). In the sequel this parameter will be varied, jointly 
with the ratio $a/c$ that characterises helicity.  

The parameter $R_0$ 
which determines the range of the attractive interaction, shall have the following value
$
R_0  \ = \ 5.0 \ \ {\rm \AA }
$
throughout;  this choice is in line with 
all-atom molecular dynamics simulations where any long range interactions between atoms 
are commonly cut off, sharply, beyond distances around 10 \AA ngstr\"om or so.

The attractive interaction can be given a physical interpretation, in terms of  ``hydrophobic" forces:
In the case of {\it e.g.} a protein chain under physiological conditions, there is an effective 
attractive interaction between those amino acids 
which are considered ``hydrophobic" \cite{degennes-1979,grosberg-1994,schafer-1999}.
Thus in the presence of the attractive interaction (\ref{U}) the energy
function (\ref{hamiltonian}) models a chain made of ``hydrophobic" residues, with ``hydrophobicity" that depends
on the value of $U_0$.

It is noted that qualitatively,  the present 
results have been found to be quite insensitive to the details of the profile of the potential $U(r)$.  
Accordingly (\ref{U}) can be considered "universal".

\subsection{Monte Carlo Algorithms}
\label{sect:Simulation}

The protein folding problem \cite{freddolino-2010,dill-2012,piana-2014}
is notorious for its computational complexity.
A comprehensive  all-atom simulation with {\it Anton} \cite{shaw-2008,shaw-2009} which is by far the fastest molecular dynamics machine
available, can produce no more than 
a few micro seconds of a folding trajectory per day {\it in silico}, in the case of proteins with less than 100 amino
acids. Since many proteins take seconds, even days to fold into their native state starting from an initial random
conformation, it could take thousands of years to fold such a protein with presently
available computer resources. Even in the case of effective off-lattice models such as (\ref{hamiltonian}), (\ref{fvol})
the simulation of a full folding trajectory is a formidable computational 
challenge, even with the most powerful computers available.
Accordingly, to identify an effective computational scenario, the performance of three different Markovian 
Monte Carlo algorithms \cite{berg-2004} have been tested.  The aim has been to  identify a 
Monte Carlo algorithm that has  the fastest rate of convergence towards a 
thermal equilibrium state, in the case of a single polymer chain. 
 In these tests {\it only} the  forbidden volume constraint (\ref{fvol}) 
has been used, the effect of the attractive (``hydrophobic") interaction has not been included.
 The three algorithms are

\vskip 0.2cm

1) Heat Bath algorithm 

2) Metropolis algorithm

3) Mixed algorithm

\vskip 0.2cm

The curvature and torsion angles 
are updated according to a probability distribution,  that satisfies the 
detailed balance condition
\[
P(\{ \kappa_{\text{new}}, \tau_{\text{new}} \}, \{ \kappa_{\text{old}}, \tau_{\text{old}} \} ) \exp ( - \beta H (\{ \kappa_{\text{old}}, \tau_{\text{old}} \} ) ) 
\]
\[
= P(\{ \kappa_{\text{old}}, \tau_{\text{old}} \}, \{ \kappa_{\text{new}}, \tau_{\text{new}} \} ) \exp ( - \beta H (\{ \kappa_{\text{new}}, \tau_{\text{new}} \} ) )
\]
Here $\beta$ is the inverse Monte Carlo temperature.  The equilibrium distribution 
\begin{equation}
\exp \! \left\{ - \beta H (\kappa, \tau  )  \right\}
\label{weight}
\end{equation}
of a canonical ensemble is obtained in the limit of an infinite  number of updates. 
Each update consists of a ``walk" through the entire chain with a provisional revision
of each value ($\kappa_i , \tau_i$) which is subject to the requirement that the forbidden
volume constraint (\ref{fvol}) is preserved ; the three algorithms  differ from each
other only in the manner how the  new values ($\kappa_i^{\text{new}}, \tau_i^{\text{new}}$) are generated.

It should be kept in mind in the sequel,  
that the Monte Carlo temperature $T = \beta^{-1}$ is {\it not} equal to the physical temperature
factor $k_B\theta $ where  $k_B$ is the Boltzmann constant and the temperature $\theta$ 
is measured in Kelvin scale. $T$ is dimensionless quantity like energy $H$ (see eq. (\ref{hamiltonian}) ) and all parameters in it (see Table \ref{table-1}).  Instead, in the low temperature collapsed regime general renormalisation group 
arguments \cite{krokhotin-2013} propose that dimensionless Monte Carlo temperature $T$ is connected with real physical temperature in the following way:
\begin{equation}
\ln T \ =  \  k_B \theta + \dots
\label{Boltz}
\end{equation}

\subsubsection{Heat Bath algorithm}
In the Heat Bath algorithm, new values ($\kappa_i^{\text{new}}, \tau_i^{\text{new}}$)  are generated randomly,
according to probability  distributions
\begin{equation}
P(\kappa_i^{\text{new}}) \ = \ \frac{1}{Z_{i, \kappa}} \exp\left\{ -\beta H_{i,
\kappa}(\kappa_i^{\text{new}})\right\}
\label{kappa_distr}
\end{equation}
and
\begin{equation}
P(\tau_i^{\text{new}}) \ = \ \frac{1}{Z_{i, \tau}} \exp\left\{ -\beta H_{i, \tau}(\tau_i^{\text{new}})\right\}
\label{tau_distr}
\end{equation}
Here $H_{i, \kappa}$  and  $H_{i, \tau}$ are  the sum of all those terms in the Hamiltonian (\ref{hamiltonian}) that
contain $\kappa_i$ and $\tau_i$, respectively, with the given index $i$. The  
$Z_{i, \kappa}$ and $Z_{i, \tau}$  are
normalisation factors. The updated values of $\kappa_i^{\text{new}}$ and $\tau_i^{\text{new}}$ do not
depend on the previous values of $\kappa_i$ and $\tau_i$.

The probability density for $\kappa_i^{\text{new}}$ has the form
\begin{equation}
P(\kappa_i) \sim \exp\{ -c_1 \kappa_i ^4 - c_2
\kappa_i^2 - c_3 \kappa_i \}
\label{hbkappa}
\end{equation}
where
\begin{eqnarray}
 c_1 & = & \beta  q  \nonumber \\ 
c_2 & = & \beta  (2 - 2 q m^2 + \frac{c}{2}  d \tau_i^2 -a b \tau_i)  \\
c_3 & = & \beta (-2 (\kappa_{i+1} +\kappa_{i-1} ) ) \nonumber
\end{eqnarray}
Thus  (\ref{hbkappa}) is non-Gaussian. On the other hand, 
the probability density $P(\tau_i^{\text{new}})$ has the  
Gaussian profile
\begin{equation}
P(\tau_i) \sim \exp \{ -\beta (\frac{c}{2} [d \kappa_i^2+1] \tau_i^2
-a  \tau_i) \}
\label{hbtau}
\end{equation}
Rejection sampling has been used to generate random
numbers according to these probability distribution:
After generating  $\kappa_i^{\text{new}}$ and $\tau_i^{\text{new}}$  the forbidden volume 
condition (\ref{fvol}) is checked, and the update is rejected when the condition is violated.

\subsubsection{Metropolis algorithm}
New values of $\kappa$ and $\tau$ are generated according to
Gaussian probability distributions, which is centered  at the old values. 
The dispersion of each Gaussian
can be adjusted, to enhance the convergence of the algorithm. 
The new values of $\kappa$ and $\tau$  are  accepted or rejected,  in the same manner as in the 
conventional Metropolis algorithm. For example, in the case of $\tau_i$ the
probability of acceptance of a new value  is
\begin{equation}
P(\tau_i)=\min\{1, \exp(-\beta \Delta H) \}
\end{equation}
where $\Delta H$ is the difference of the energy between the new and the old
configurations. In addition, the self-avoidance condition  (\ref{fvol})  is also verified at each step.

\subsubsection{Mixed algorithm}
The values of $\kappa_i^{\text{new}}$  are generated in the same manner as in the Heat Bath algorithm, while 
for  $\tau_i^{\text{new}}$ the Metropolis
algorithm is used. The convergence of the algorithm can be adjusted, by changing the
dispersion of the Gaussian distribution in the $\tau$ update.

\subsubsection{Algorithm comparison}
\label{sect:comparison}

{\it A priori}, each of the three Monte Carlo algorithms  should converge towards  the same
equilibrium distribution albeit at a different speed. Thus, when the  Markovian length 
is not sufficient and the equilibrium distribution is not yet reached, the 
result depends on the algorithm and the number of Monte Carlo steps. 
Accordingly, the three algorithms have been tested and compared, to identify the one with the fastest 
convergence rate towards a known equilibrium  state; for this, the zero temperature ground 
state  described in sub-section  II. E. is utilised.  In these tests, the  following parameter 
values have been used  in the hamiltonian (\ref{hamiltonian}),
\begin{equation}
\begin{matrix} a &=& \! \! -1.0 \times 10^{-4}  \\
c &=&  ~ 1.0 \times 10 ^ {-4}  
\end{matrix}
\label{paras1}
\end{equation}
These values have been  chosen to reproduce a  monotonous $\alpha$-helical structure 
in a manner which is consistent with the forbidden volume constraint (\ref{fvol}),
as the lowest energy conformation; the choice is not unique. The simulations 
have been performed with varying chain lengths, from  N=100 to  N=900.
In each case, simulated annealing has been used and the initial configuration is always 
a linear straight rod with 
$
\kappa_i \ = \ \tau_i \ = \ 0
$.
The initial configuration is first heated to very high temperature values  (up to $T_{max}=100$), where the 
structure is fully thermally randomised. This is followed by a slow cooling period, to the 
target temperature. The cooling  takes place with small temperature steps, with each step 
equal to $\Delta T = 0.05 \ - \ 
 0.5$ in logarithmic scale. 
After  each  step,  $10^3 - 5 \times 10^4$ Monte Carlo updates are  performed along the whole chain, 
to ensure that it becomes thermalized to the ambient temperature. Here the term  ``termalized'' means that the Marcovian chain reaches its equilibrium distribution. We will call  ``thermalization length'' the number of Monte Carlo updates per one step of cooling process.
 The longer the chain, the longer the thermalization. 
For each temperature step only the last configuration is used for calculation of 
observables. Thus during one cooling process one final (thermalized) chain configuration is obtained
for each temperature value.
The final phase diagrams have been calculated using 
5000 Monte Carlo updates per one step of cooling, equal to $\Delta T=0.05$ in logarithmic scale. 
One single cooling procedure takes  6 CPU-hours, producing one configuration for each temperature value. 
Final statistics is compiled from 128 thermalized chain configurations, for each set of parameters values.

\begin{figure}
\centering
{\begin{minipage}[t]{.5\textwidth}
  \raggedright
  \includegraphics[trim = 0mm 75mm 0mm 70mm, width=1\textwidth,  angle=0]{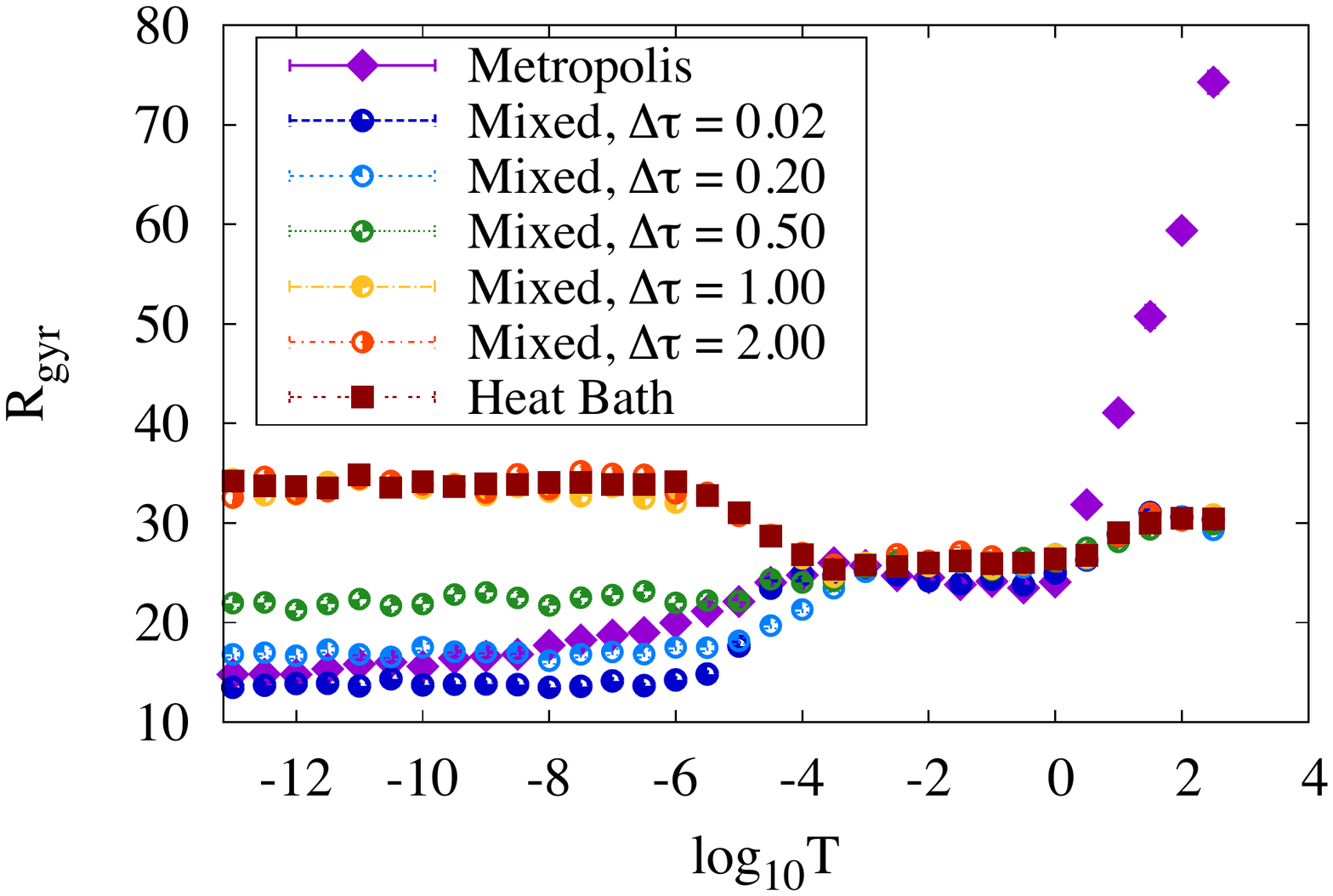}
  \caption{(Color online) Dependence of $R_{gyr}$ on temperature for  identical 
  homopolymer chains simulated with different algorithms. Here $\Delta \tau$ is dispersion of 
  Gaussian distribution for generation of $\tau_{\text{new}}$ in mixed algorithm. 
  In conventional  Metropolis algorithm the dispersions  are $\Delta \kappa = \Delta \tau=0.01$. Length of polymer chain is $N$=100, parameters of the Hamiltonian are the ones from Table I.}
  \label{fig-1}
\end{minipage}}
\end{figure}
\hfill
\noindent

\begin{figure}
\centering
{\begin{minipage}[t]{.5\textwidth}
  \raggedright
  \includegraphics[trim = 0mm 75mm 0mm 70mm, width=.9\textwidth,  angle=0]{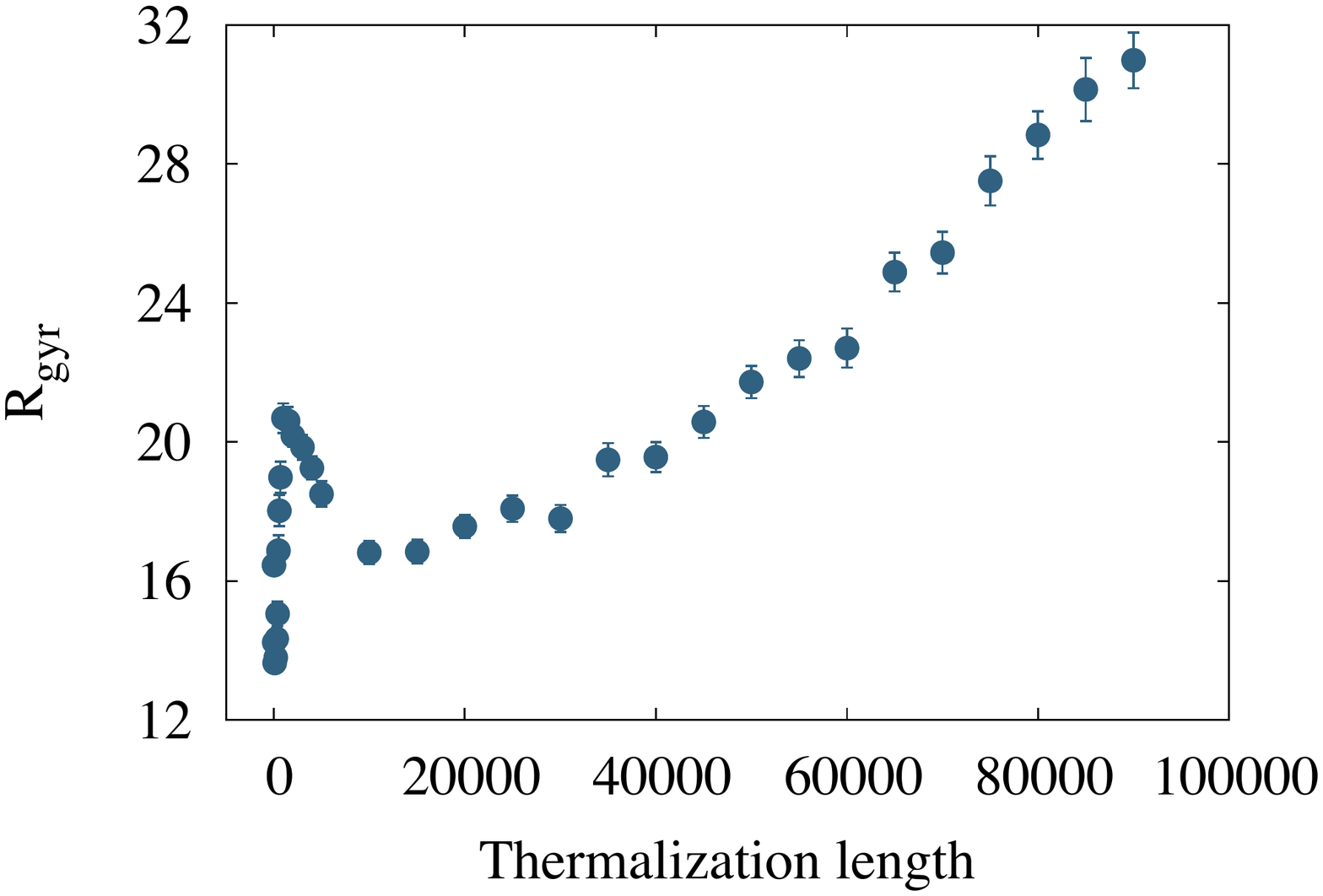}
  \caption{Dependence of $R_{gyr}$ on the thermalization length in Metropolis algorithm. 
  The number of updates, per one step in simulated annealing process (thermalization length), is shown along the horizontal axis. Temperature is $T=10^{-12}$, length of polymer $N$=100, and parameters of Hamiltonian are taken from the Table I.}
  \label{fig-2}
\end{minipage}}
\end{figure}

\begin{figure}
\centering
{\begin{minipage}[t]{.5\textwidth}
  \raggedright
  \includegraphics[trim = 0mm 75mm 0mm 70mm, width=.9\textwidth,  angle=0]{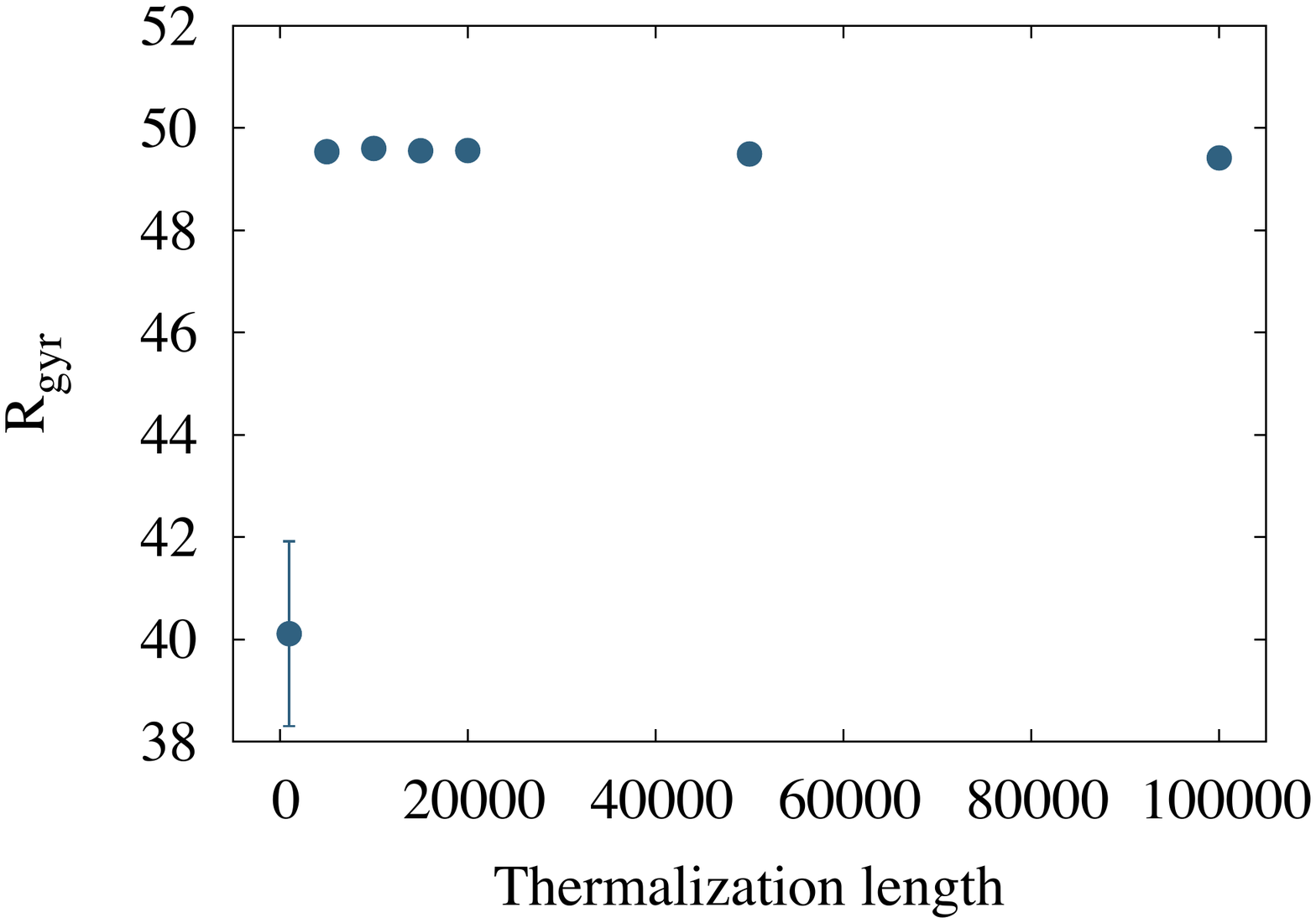}
  \caption{Dependence of $R_{gyr}$ on the thermalization length in Mixed algorithm. 
  The number of updates, per one step in simulated annealing process (thermalization length), is shown along the horizontal axis. Parameters of this run are the same as for the figure 2.}
  \label{fig-2a}
\end{minipage}}
\end{figure}

In the sequel the radius of gyration (\ref{Rg})  is utilised 
as the principal observable,  to characterise the geometry of the homopolymer chain. 
Its value is calculated  as an average over statistically independent chain configurations, 
produced during a Monte-Carlo process.  
In Figure  \ref{fig-1}  the value of the radius of gyration (\ref{Rg}) is compared, as a function of Monte Carlo temperature 
for a chain  with N=100 monomers. The results obtained  using the Metropolis algorithm are found to be very
different from those obtained using the Heat Bath and Mixed algorithms; for  small dispersion
$\Delta \tau$ the Mixed algorithm coincides with the Metropolis algorithm.  But when  $ \Delta \tau$ increases, 
the Mixed algorithm  approaches the Heat Bath algorithm as shown in the Figure. 

There is no {\it a priori}  reason why the results for the three algorithms should be different:  The stationary 
distribution is the same, in each of the  three algorithms.  But it is found that the Metropolis algorithm 
converges  {\it very} slowly towards the equilibrium distribution. For this, the dependence of the result on 
the length of simulation has been analysed. In the case of the Metropolis algorithm, the 
results are shown in the  Figure \ref{fig-2},  for  $T= 10^{-13}$  which is the lowest  
Monte Carlo temperature value that has been used in the present  simulations. As shown in the Figure,
$R_{gyr}$ continues to increase with increasing length of thermalization. 
The  Metropolis algorithm  approaches the  Heat Bath algorithm very slowly, as a function of the 
simulation time and even at the present, relatively long simulation times the chain is still far from equilibrium distribution. 

On the other hand, the results shown in Figure \ref{fig-2a} demonstrate that in case of the mixed algorithm the
radius of gyration $R_{gyr}$ approaches a fixed value with increasing of the thermalization length. It is concluded
that using the present simulation times, the Markovian homopolymer reaches a stationary distribution.
  Either the Heat Bath algorithm, or alternatively the Mixed algorithm with sufficiently 
large $\Delta \tau$, should be used to try and describe the thermal equilibrium configurations. 

\begin{figure}
\centering
{\begin{minipage}[t]{.5\textwidth}
  \raggedright
  \includegraphics[trim = 0mm 75mm 0mm 70mm, width=1\textwidth,  angle=0]{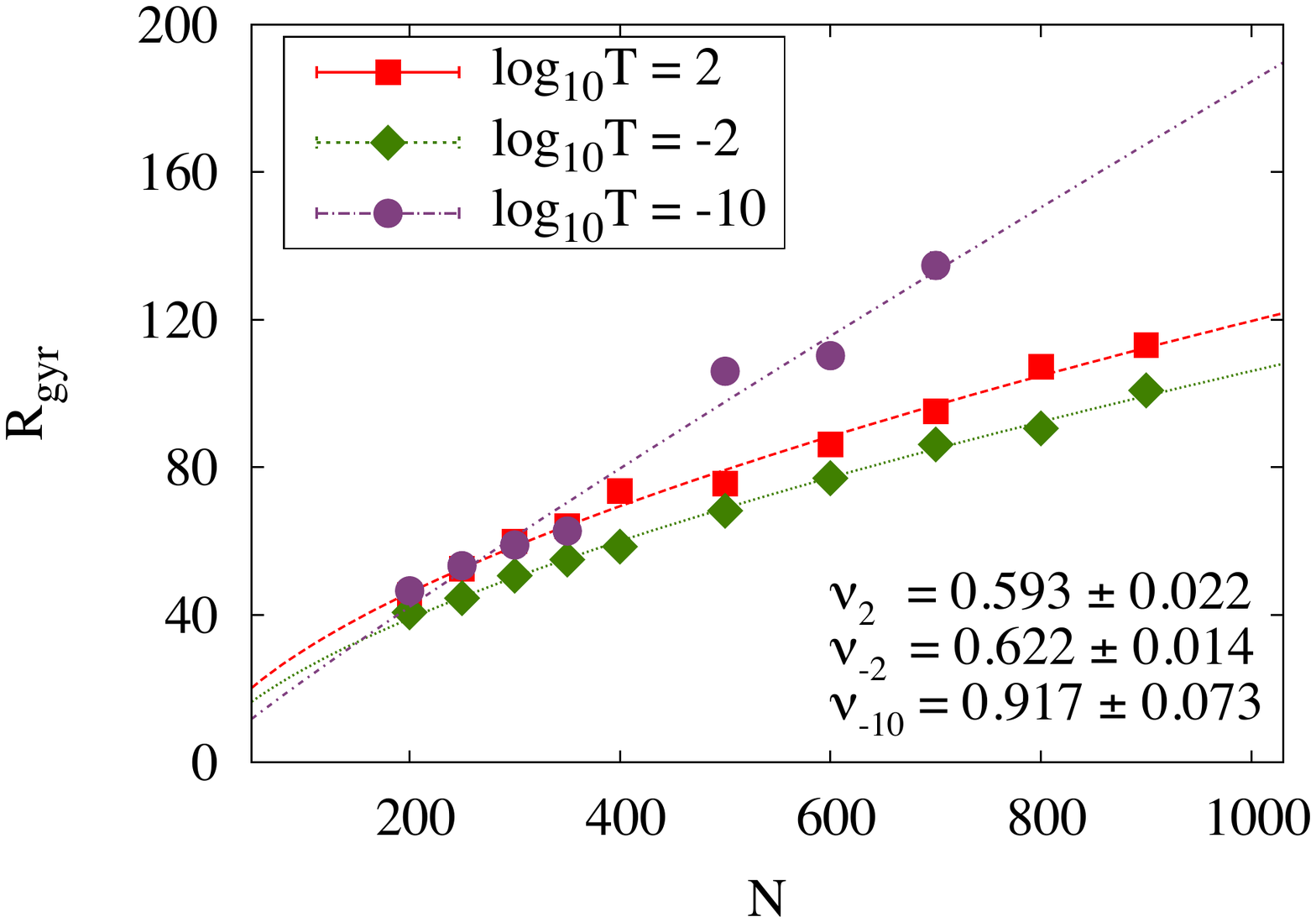}
  \caption{(Color online) Compactness index $\nu$  at different temperatures for a homopolymer chain. Simulation performed with Heat Bath algorithm. Parameters of Hamiltonian are taken from the Table I.}
  \label{fig-3}
\end{minipage}}
\end{figure}

The  compactness index $\nu$ {\it i.e.} the inverse of the Hausdorff dimension 
has  also been  inspected, using the three different algorithms. The results for the Heat Bath algorithm  
are shown in Figure \ref{fig-3}. 
%
Between $T=100$ and $T=1000$ the value of $\nu$ is essentially  
temperature independent, and   apparently corresponds to the SAW phase. 
At very low temperatures a  transition to the rigid rod state, with $\nu \approx 1$, is observed. 
This result is in line with the general arguments that are presented in 
sub-section \ref{sect:attractive},  on the expected phase structure of the homopolymer 
model  (\ref{hamiltonian}), (\ref{fvol}). 

\subsubsection{The final algorithm}
\label{sect:algorithm}

On the basis of the results from the  test runs, the following improved Heat Bath algorithm is employed in the sequel:  
The final probability distribution is 
\begin{equation}
P=\frac {1} {Z} \exp\{-\beta [ H +\sum_{i<j} U (\vec r_i -\vec r_j )]\}
\label{distribution}
\end{equation}
Here $H$ is the Hamiltonian (\ref{hamiltonian}) and $U$ is the potential (\ref{U}). 
The Metropolis algorithm is used for acceptance, but with a proposal distribution that coincides with the
Heat Bath algorithm: The
new values of $\kappa_i$ and $\tau_i$ are generated using the 
distributions (\ref{kappa_distr}) and (\ref{tau_distr}).  The ensuing
homopolymer configuration is then accepted, provided it  satisfies both 
the self-avoidance condition (\ref{fvol}) and the Metropolis accept-reject condition 
that utilises the residual energy 
\begin{equation}
E_U=\sum_{i<j} U (\vec r_i -\vec r_j )
\label{E_U}
\end{equation}
The acceptance criterion is
\begin{equation}
\exp (-\beta \Delta E_U) > \lambda
\label{acceptance}
\end{equation}
where $\lambda$ is a random number which is uniformly distributed between 0 and 1, and $\Delta E_U$ is the 
change in $E_U$  under the update of $\kappa_i$ and $\tau_i$.

\begin{figure}
  \centering
  {\begin{minipage}[t]{.5\textwidth}
  \raggedright
  \includegraphics[trim = 0mm 75mm 0mm 70mm, width=1\textwidth,  angle=0]{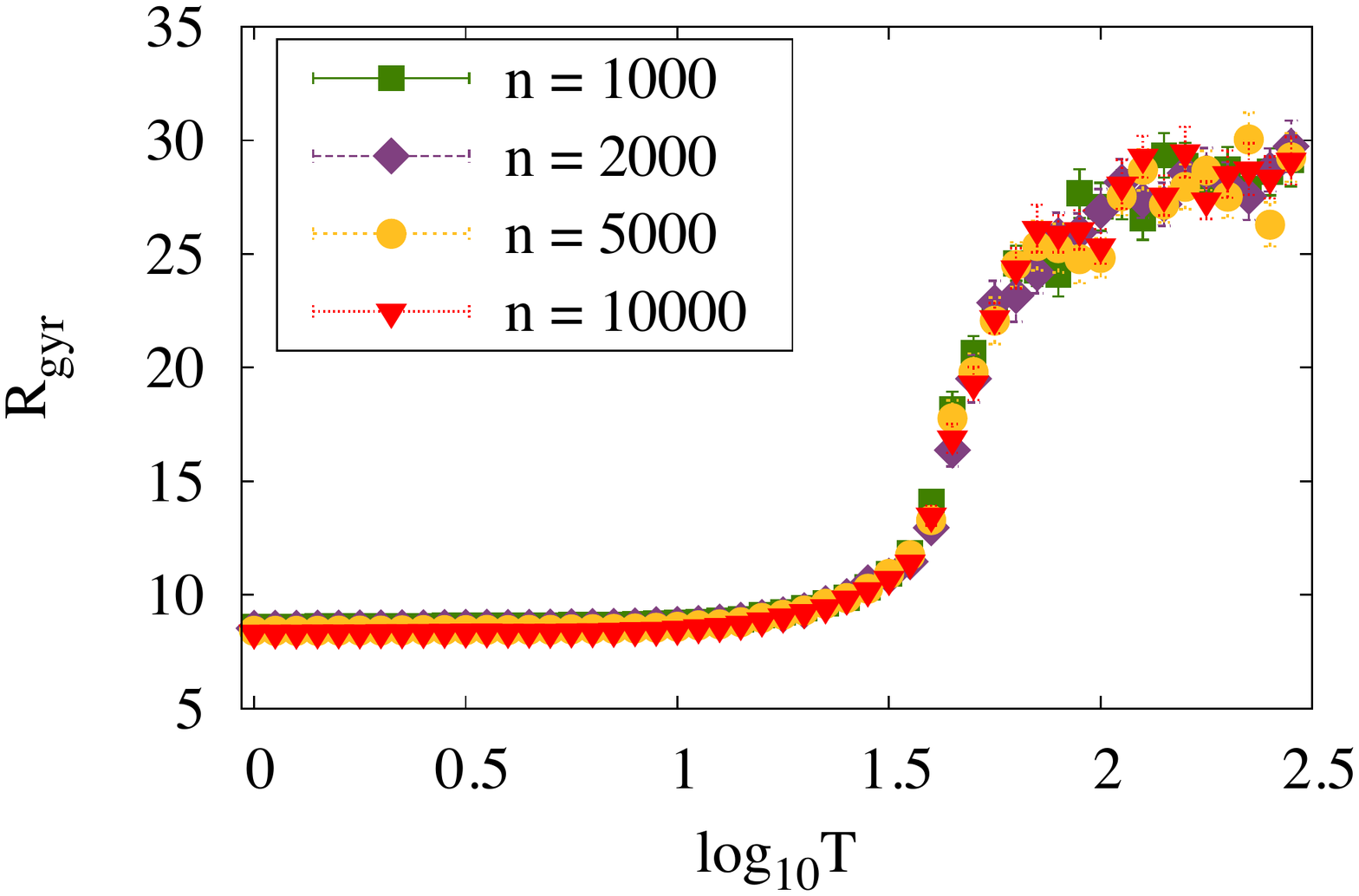}
  \caption{(Color online) Dependence of $R_{gyr}$ on temperature T for a  chain with 100 monomers, in the truncated model. 
  Here $n$ is the number of Monte Carlo updates per one simulated annealing step in temperature (thermalization length). The final algorithm described in paragraph H5 of the section II is used.}
  \label{fig-4}
\end{minipage}}
\end{figure}

Finally, the algorithm  has been calibrated by considering the  limit of a {\it truncated}  Hamiltonian , 
where the Hamiltonian $H$ is removed and only the attractive potential (\ref{U})
together with the forbidden volume constraint (\ref{fvol}) are retained. The
numerical value of $U_0$ determines solely the scale of the Monte Carlo temperature  $T$, in the present 
simulations the value $U_0=15$  is used. The results are shown in Figure \ref{fig-4}.
A smooth transition is observed in the radius of gyration, from larger values  at high temperatures to 
smaller values at low temperatures. 

\begin{figure}
  \centering
{\begin{minipage}[t]{.5\textwidth}
  \raggedright
  \includegraphics[trim = 0mm 75mm 0mm 70mm, width=1\textwidth,  angle=0]{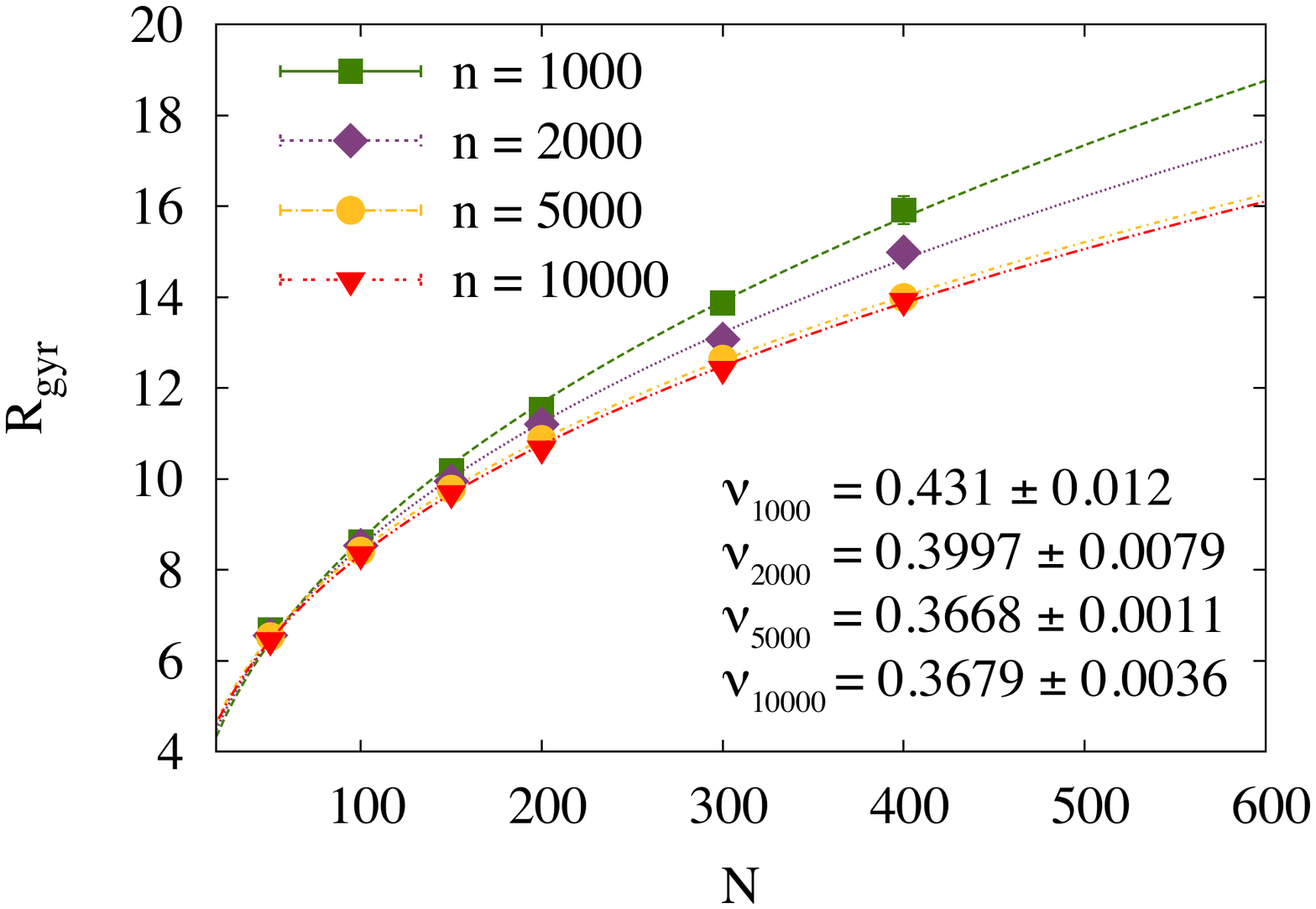}
  \caption{(Color online) Dependence of $R_{gyr}$ on the length of polymer chain  $N$ for different thermalization lengths with $T= 1$, in the truncated model. The final algorithm is used.}
  \label{fig-5}
\end{minipage}}
\end{figure}

\begin{figure}
  \centering
{\begin{minipage}[t]{.5\textwidth}
  \raggedright
  \includegraphics[trim = 0mm 75mm 0mm 70mm, width=.9\textwidth,  angle=0]{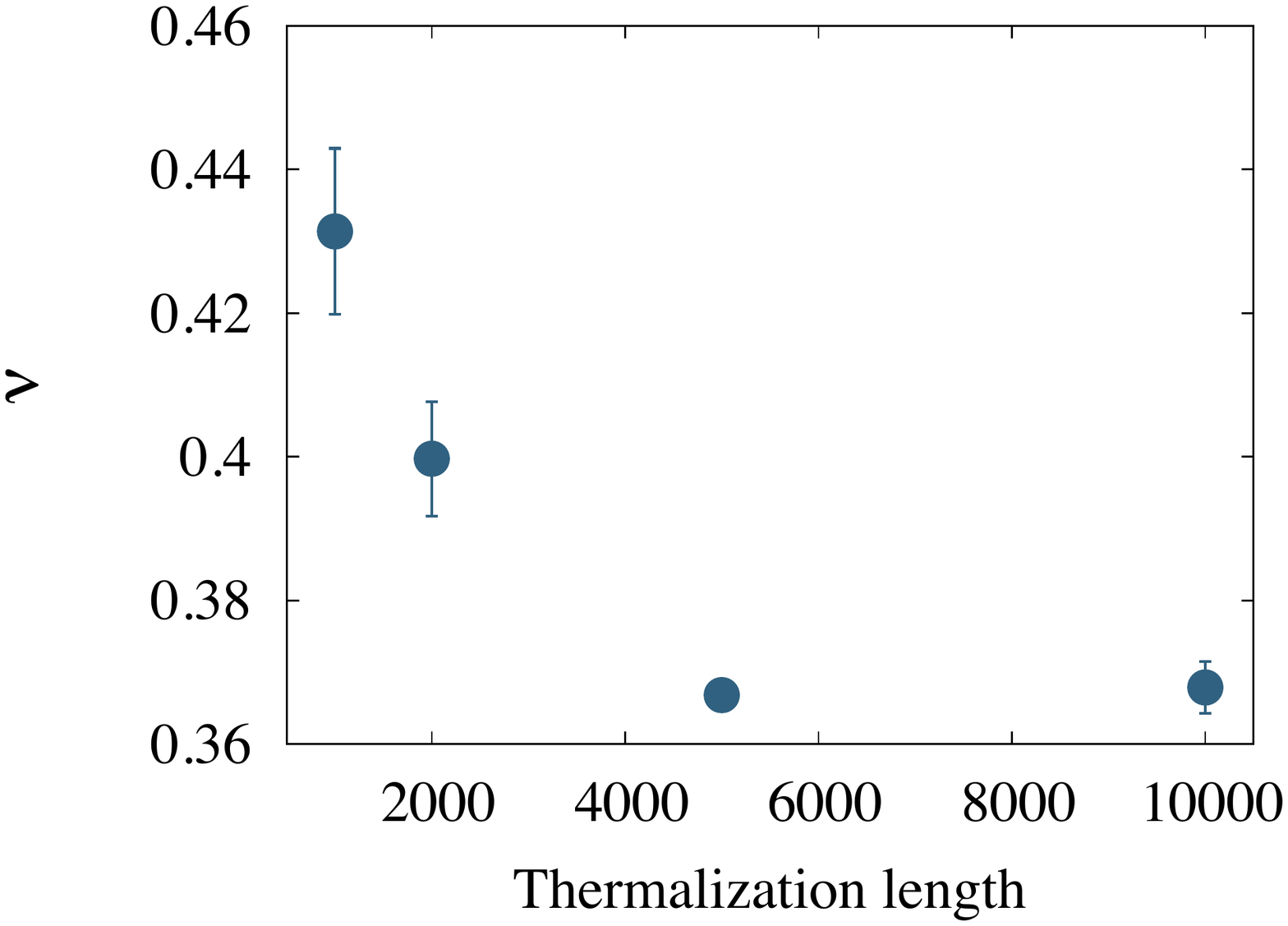}
  \caption{Dependence of compactness index $\nu$  on thermalization length when $T=1$, in the truncated model.  }
  \label{fig-6}
\end{minipage}}
\end{figure}

The Figure \ref{fig-5} 
shows how  $R_{gyr}$ depends on the length of polymer chain $N$, and fitted to the leading order 
contribution in (\ref{R}). The results  demonstrate how the equilibrium distribution is reached
with a sufficiently long thermalization length, even in the region of low temperatures where the convergence
is at its lowest.  In particular it is found that the value of the compactness index $\nu$ 
 converges towards the mean field value $1/3$ of the collapsed phase, as shown
in Figure \ref{fig-6}; the difference between the space filling $\nu = 1/3$ and the numerically deduced 
$\nu \approx $ 0.36 is attributed to the finite size corrections in (\ref{R}); the available computer power
does not enable an  identification of the amplitudes $R_1, ...$ or the critical exponents $\delta_1, ...$ in (\ref{R}) .


%
%
%


\section{RESULTS}

The present variant of the Heat Bath algorithm
has been used in extensive numerical simulations to  investigate the phase structure of the
homopolymer model (\ref{hamiltonian}), (\ref{U}).  
The results are summarised  in Figures \ref{fig-7}-\ref{fig-11}.

\subsection{Effect of parameters}

\subsubsection{Parameter  $U_0$}
\label{sect:Effects}

\begin{figure}
\centering
{\begin{minipage}[t]{.5\textwidth}
  \raggedright
  \includegraphics[trim = 0mm 75mm 0mm 70mm, width=1\textwidth,  angle=0]{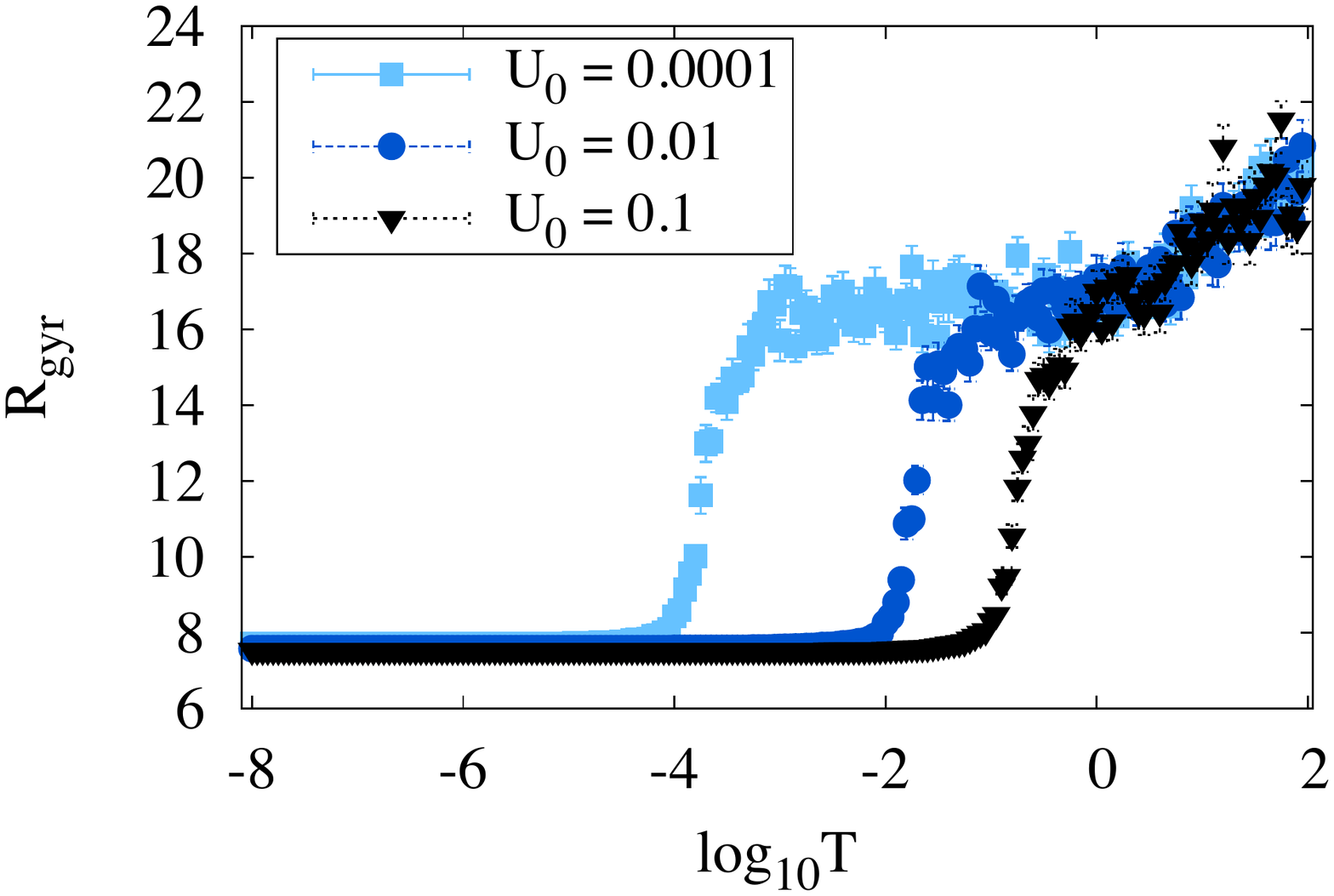}
  \caption{(Color online) Dependence of $R_{gyr}$ on temperature for a  chain with 50 monomers. Parameters of the Hamiltonian are taken from the Table I. }
  \label{fig-7}
\end{minipage}}
\end{figure}

\begin{figure}
\centering
{\begin{minipage}[t]{.5\textwidth}
  \raggedright
  \includegraphics[trim = 0mm 75mm 0mm 70mm, width=1\textwidth,  angle=0]{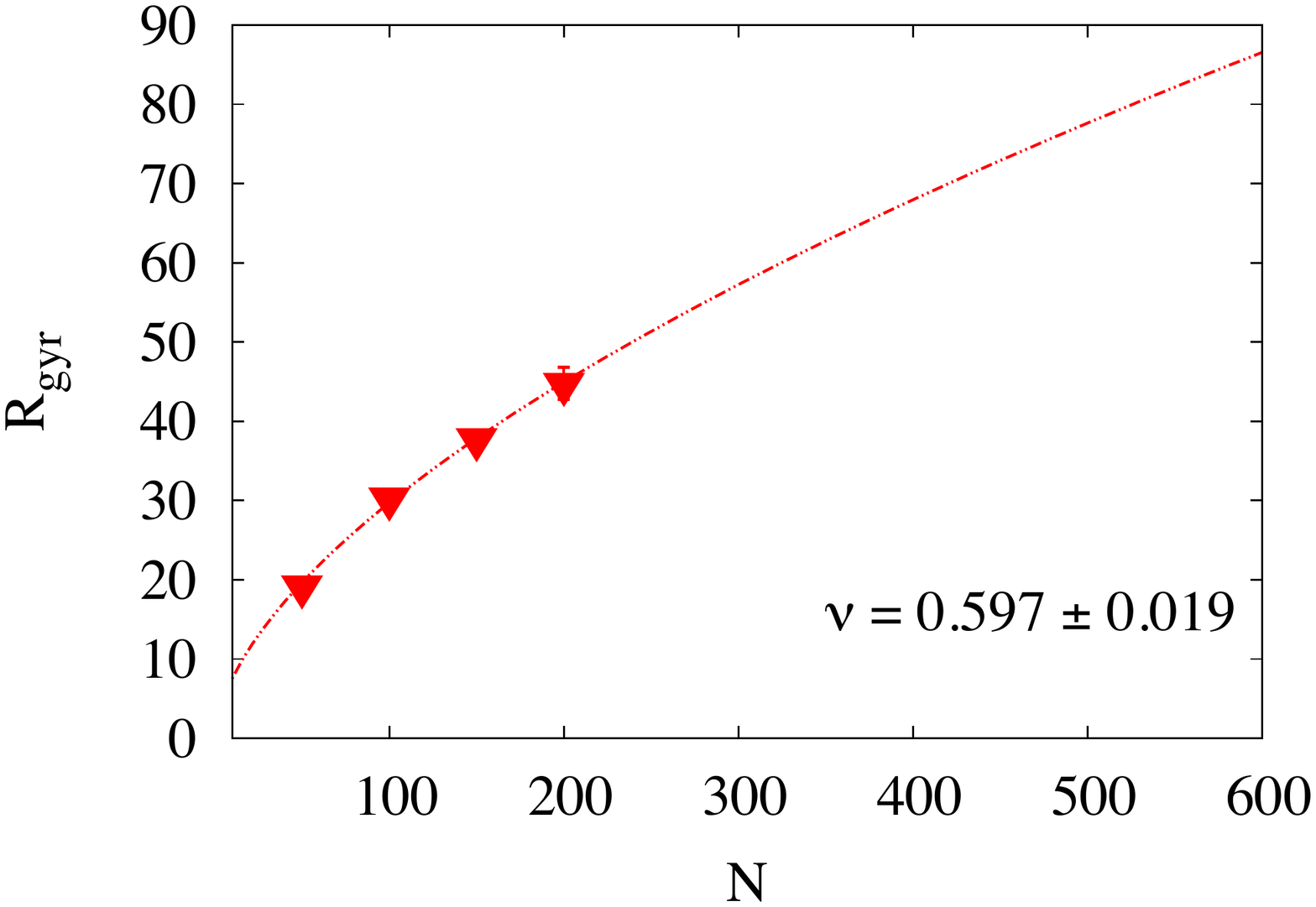}
  \caption{Dependence of $R_{gyr}$ on the length of chain for high temperature $T=10^{1.7}$ and with $U_0=0.01$. Parameters of the Hamiltonian are taken from the Table I. }
  \label{fig-8}
\end{minipage}}
\end{figure}

\begin{figure}
\centering
{\begin{minipage}[t]{.5\textwidth}
  \raggedright
  \includegraphics[trim = 0mm 75mm 0mm 70mm, width=1\textwidth,  angle=0]{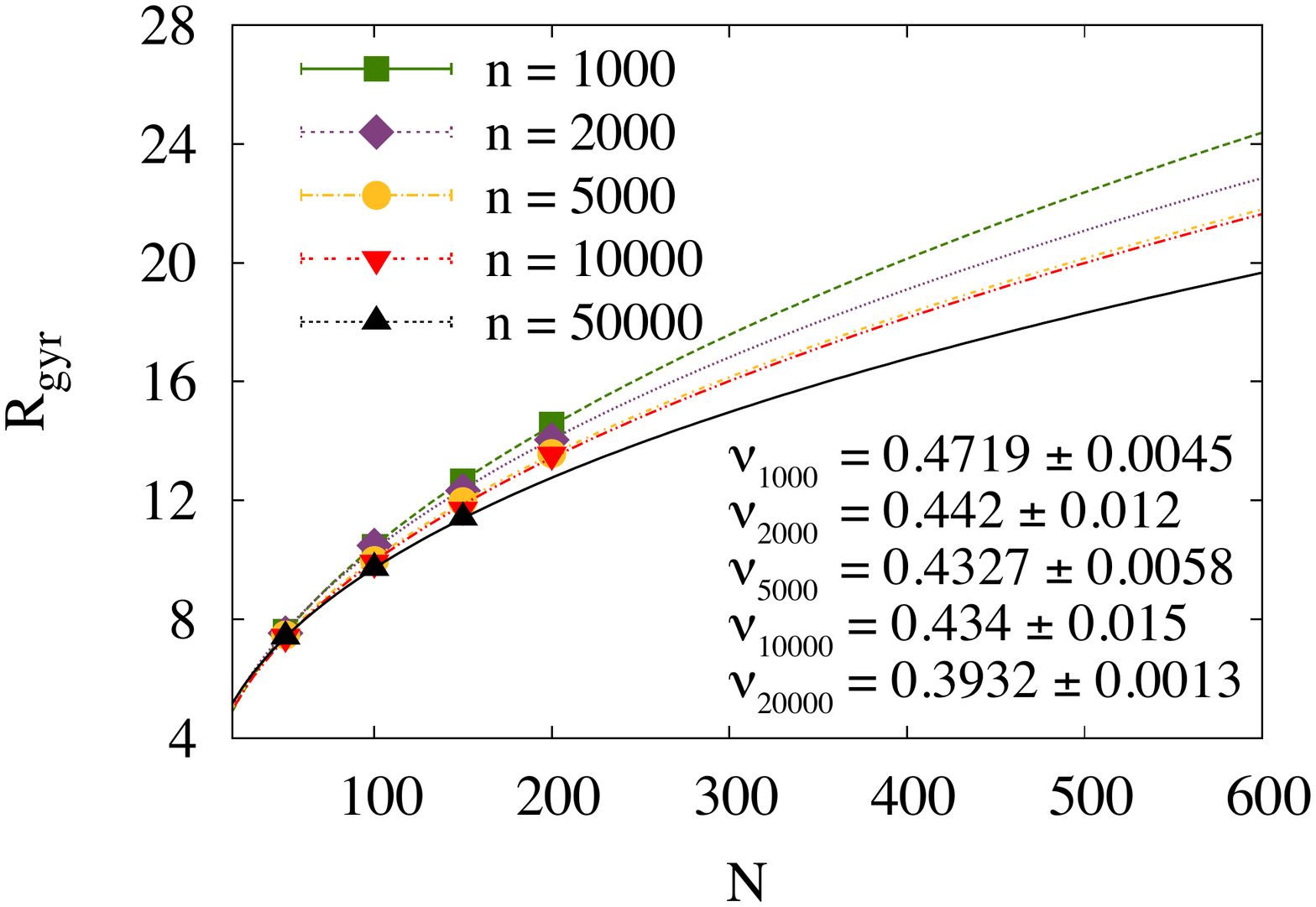}
  \caption{(Color online) Dependence of $R_{gyr}$ on the length of chain for low temperature $T=10^{-6}$ and various thermalization lengths, and with
  $U_0=0.01$. Parameters of the Hamiltonian are taken from the Table I. }
  \label{fig-9}
\end{minipage}}
\end{figure}

\begin{figure}
\centering
{\begin{minipage}[t]{.5\textwidth}
  \raggedright
  \includegraphics[trim = 0mm 75mm 0mm 70mm, width=.9\textwidth,  angle=0]{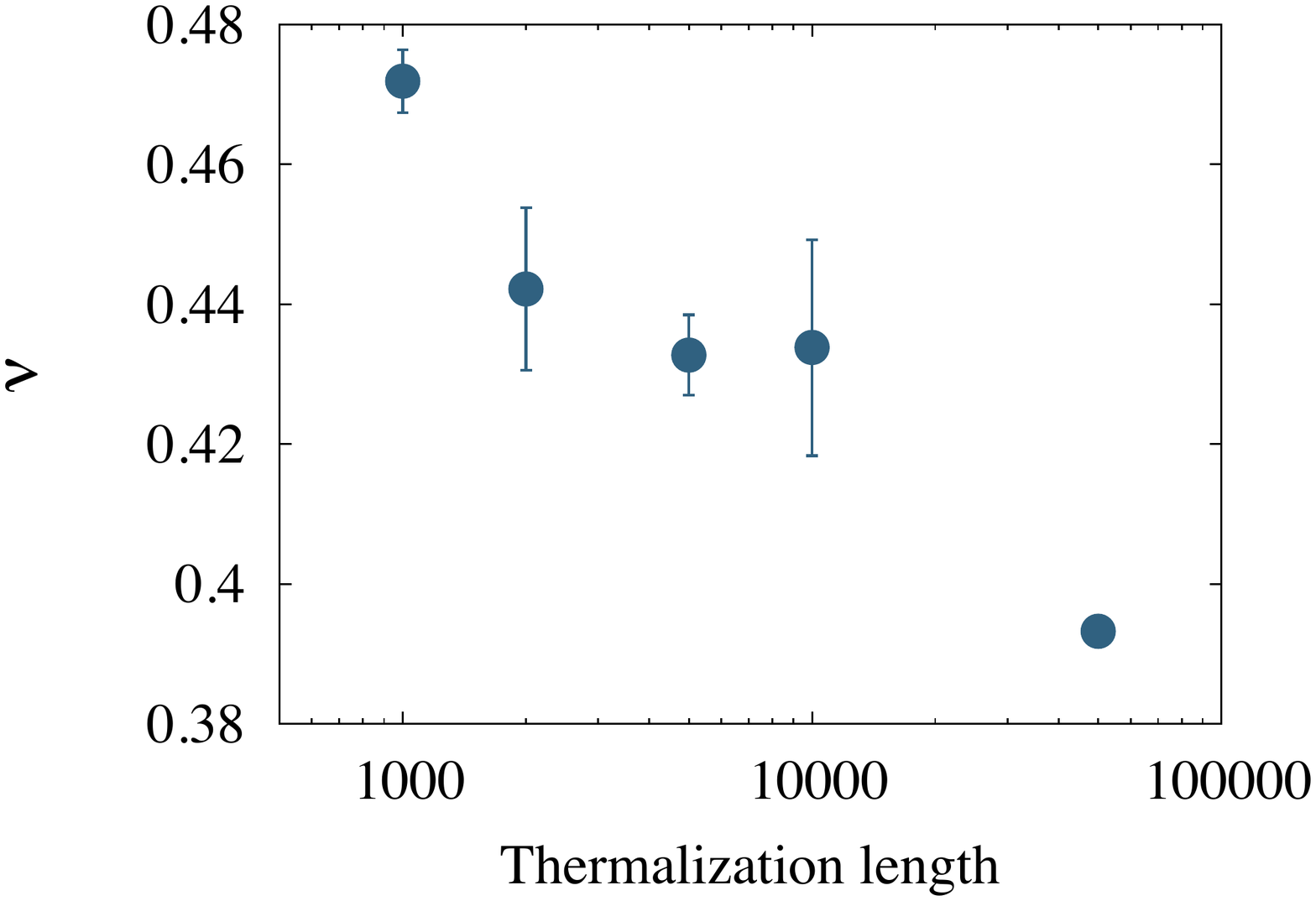}
  \caption{Dependence of compactness index $\nu$  on thermalization length. 
  Temperature is equal to $T=10^{-6}$ and $U_0=0.01$. Parameters of the Hamiltonian are taken from the Table I. }
  \label{fig-10}
\end{minipage}}

\end{figure}

Figures  \ref{fig-7}-\ref{fig-9} 
describe  the properties of the radius of gyration  for  three 
representative  values of the strength parameter $U_0$ in (\ref{U}),
\begin{equation}
U_0 \ = \ \left\{ \ \begin{matrix} 10^{-4} \\ 10^{-2} \\ 10^{-1} \end{matrix} 
\right.
\label{u_profile}
\end{equation}
For the other parameters the values (\ref{paras1}) are used.
For each  value of $U_0$, three different characteristic regimes are observed. In the high temperature 
limit the homopolymer is found in the SAW phase; this is confirmed in Figure \ref{fig-8}. 
When  the temperature decreases, the homopolymer enters a regime of decreasing 
$R_{gyr}$.  
Finally, there is the low temperature regime where the radius of gyration  $R_{gyr} $ has a  small 
value; see Figure \ref{fig-9}.  
The compactness index $\nu$ shows that when the thermalization increases the value of $\nu$ converges towards the values close to $\nu \approx 0.39$ which is indicative of the mean field value $1/3$. 
This is shown in Figure \ref{fig-10}. Again,  the 
difference between the mean field value $\nu = 1/3$ and the measured value $\nu \approx 0.36$ 
is allocated to finite size corrections in (\ref{R}); the available computer power is not sufficient to deduce the
detailed form of these corrections.

\hfill
\noindent
\hfill
\noindent
\hfill
\noindent

It is concluded that when the scale for $U_0$ exceeds that of $a$ and  $c$ the low temperature state is a collapsed
configuration.

\subsubsection{Parameters $a$ and $c$}

According to (\ref{ksolu}) the classical ground state profile of $\tau_i$ remains intact when the parameters $a$ and $c$ are changed
in such a manner that the ratio $a/c$ is constant.  To study the effect of such a change in $a$ and $c$ at finite 
temperature, 
in particular how it  conspires with the parameter $U_0$,  simulations have first been performed with 
\begin{equation}
a \ = \ c \ = \ 10^{-2} \  \ \ \ \& \ \ \ \ \ \  U_0=10^{-4}
\label{acU0}
\end{equation}
with the values (\ref{paras1}) for the remaining parameters. 
Thus, unlike in the previous simulations now the characteristic scale of the 
attractive interaction is smaller than that of the torsion
angle dependent terms in the Hamiltonian. 

\begin{figure}
\centering
{\begin{minipage}[t]{.5\textwidth}
  \raggedright
  \includegraphics[trim = 0mm 75mm 0mm 70mm, width=1\textwidth,  angle=0]{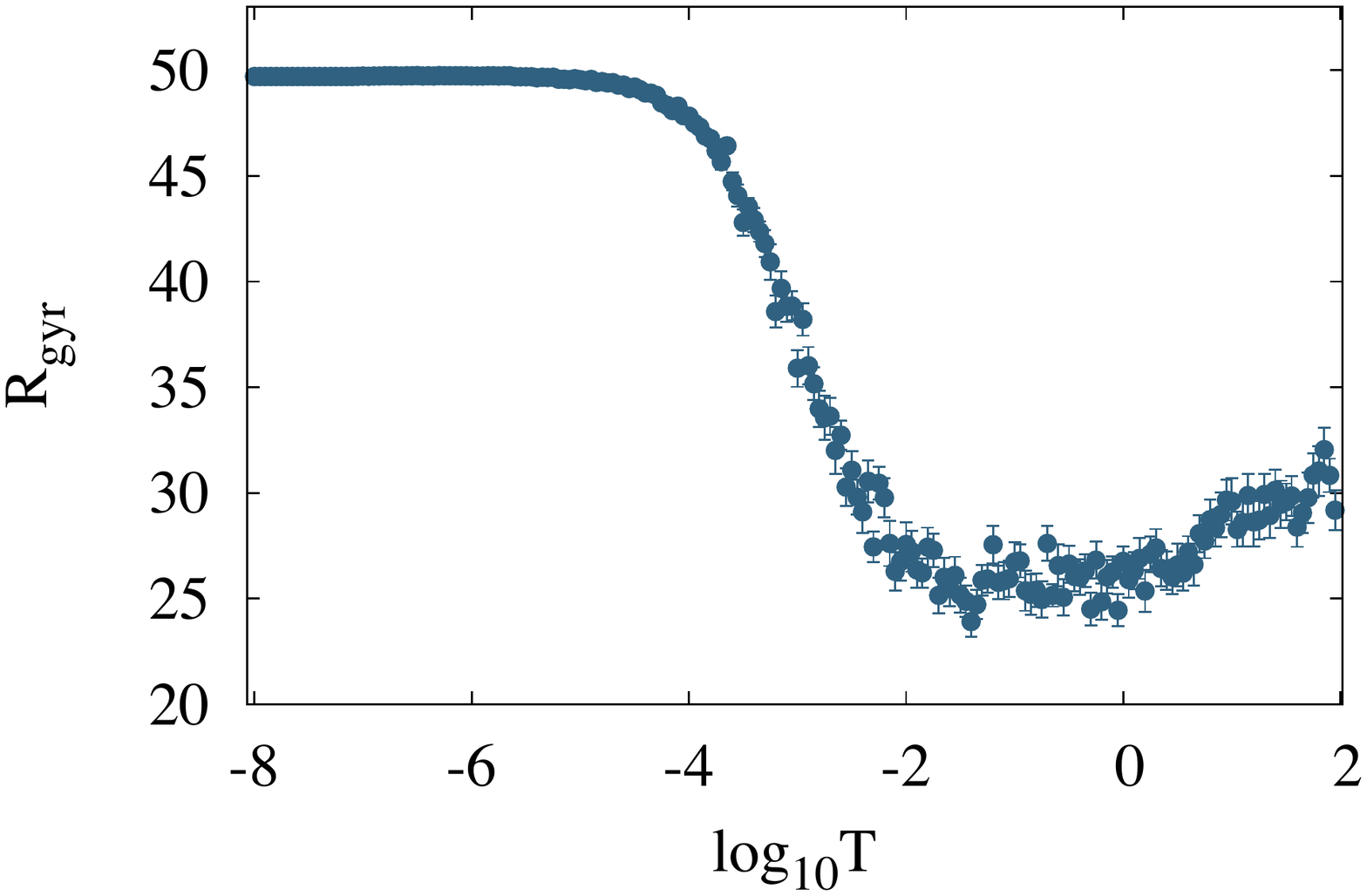}
  \caption{Dependence of $R_{gyr}$ on temperature. The chain has 100 monomers, 
  and parameters ($a,c,U_0$) are given in  (\ref{acU0}).
}
  \label{fig-11}
\end {minipage}}
\end{figure}

The results for the radius of gyration are presented in the Figure \ref{fig-11}. 
At high temperatures the chain is again in the 
SAW phase. 
Then, as temperature decreases, there is a transition to  a 
regime akin the intermediate regime shown in Figure \ref{fig-7}.
Finally, there is a low temperature regime  where 
the chain fluctuates around the classical solution (\ref{ksolu}). 
The scale of transition to the low temperature regime
is controlled by parameters $a$ and $c$ in Hamiltonian. 

It is concluded that when the scale $U_0$ of the short-range attractive interaction is smaller  
than the scale of the parameters $a$ and $c$,
the low temperature limit is described by helical structures.

\subsection{Analysis of different phase regimes}

\begin{figure}
\captionsetup[subfigure]{labelformat=empty}\captionsetup[wrapfigure]{format=plain}
\raggedright
 \subfloat[]{\includegraphics[width=0.25\textwidth]{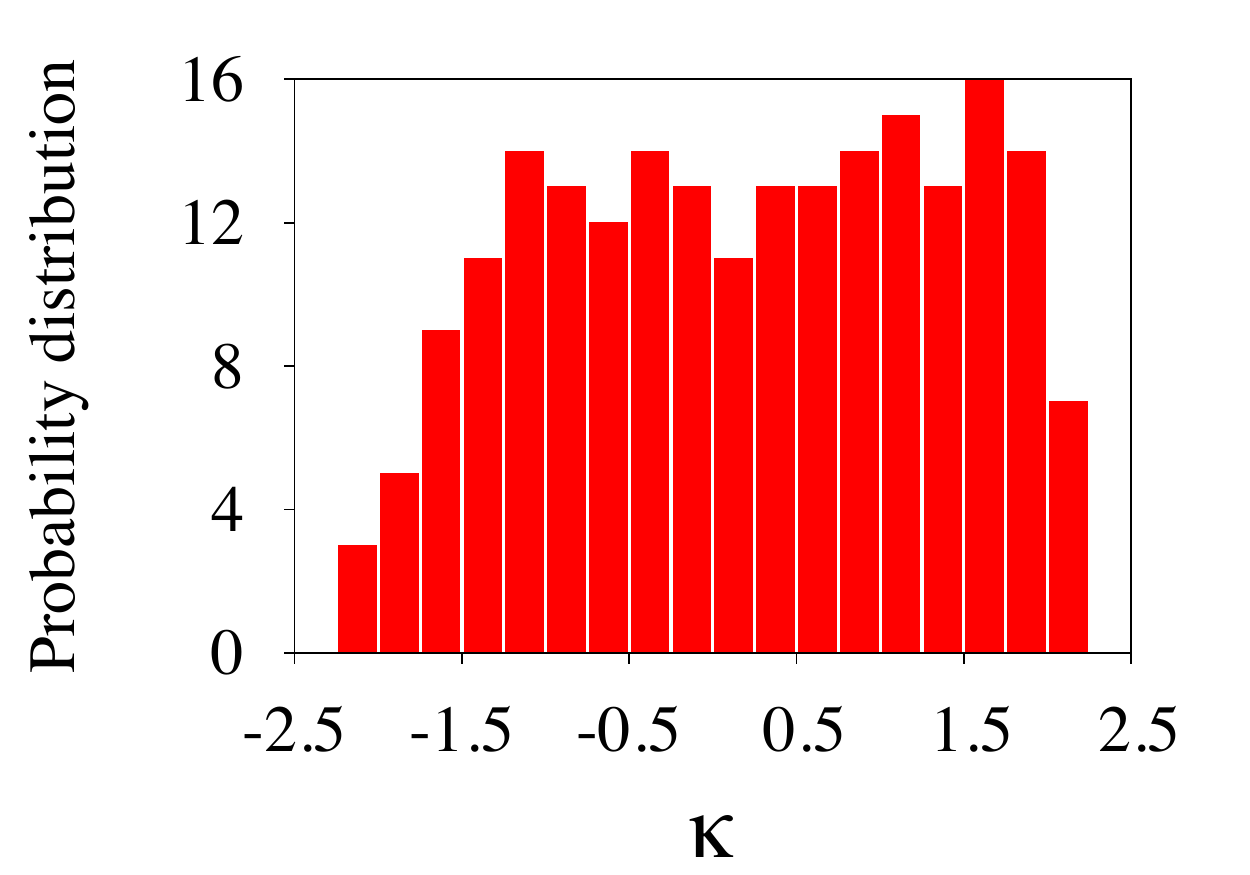}}
 \subfloat[]{\includegraphics[width=0.25\textwidth]{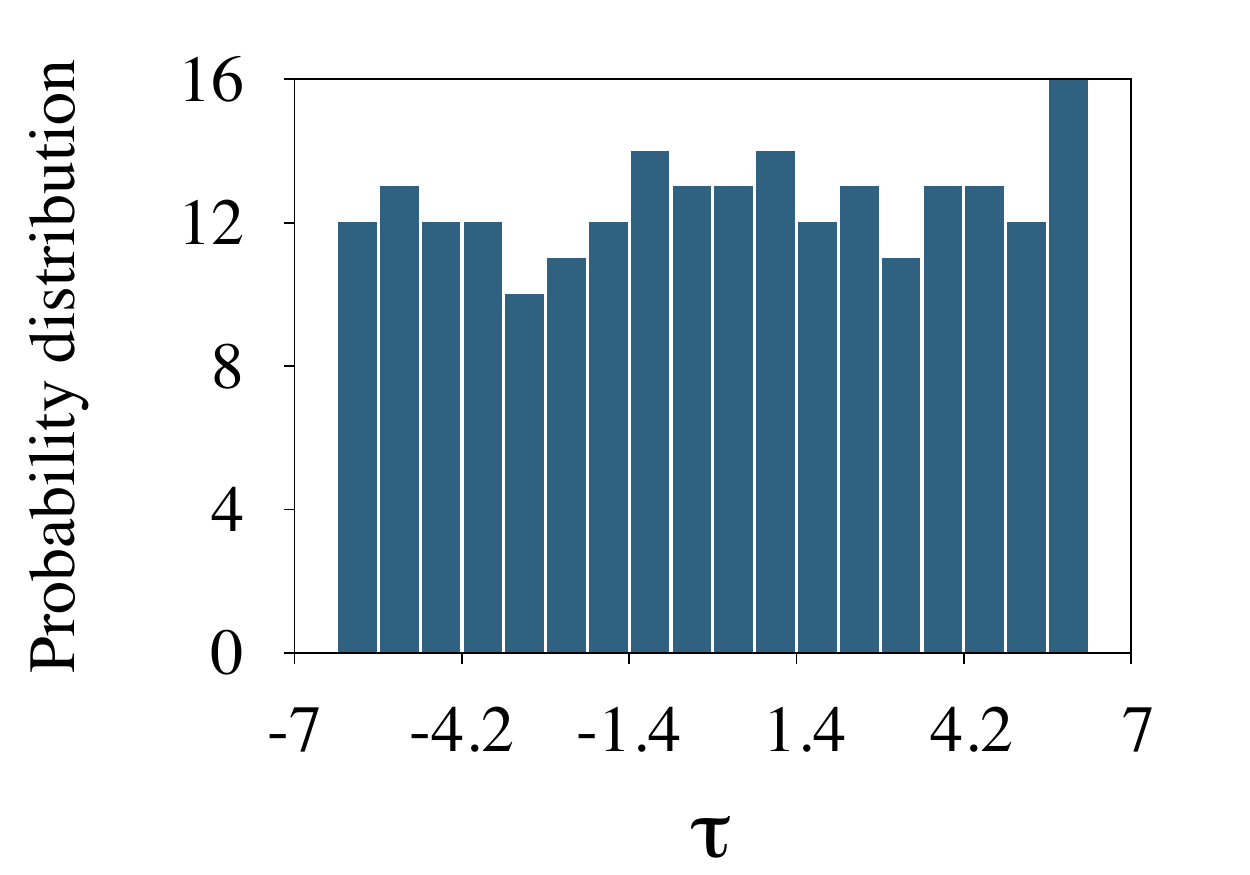}}
\vskip -0.5cm
\caption{Distribution of ($\kappa_i,\tau_i$) in the high temperature SAW phase. Simulation parameters are the same as for the figure 8 ($U_0=0.01$), temperature is $T=100$.}
\label{fig-12}
\end{figure}

\begin{figure}
\captionsetup[subfigure]{labelformat=empty}
\raggedright
 \subfloat[]{\includegraphics[width=0.25\textwidth]{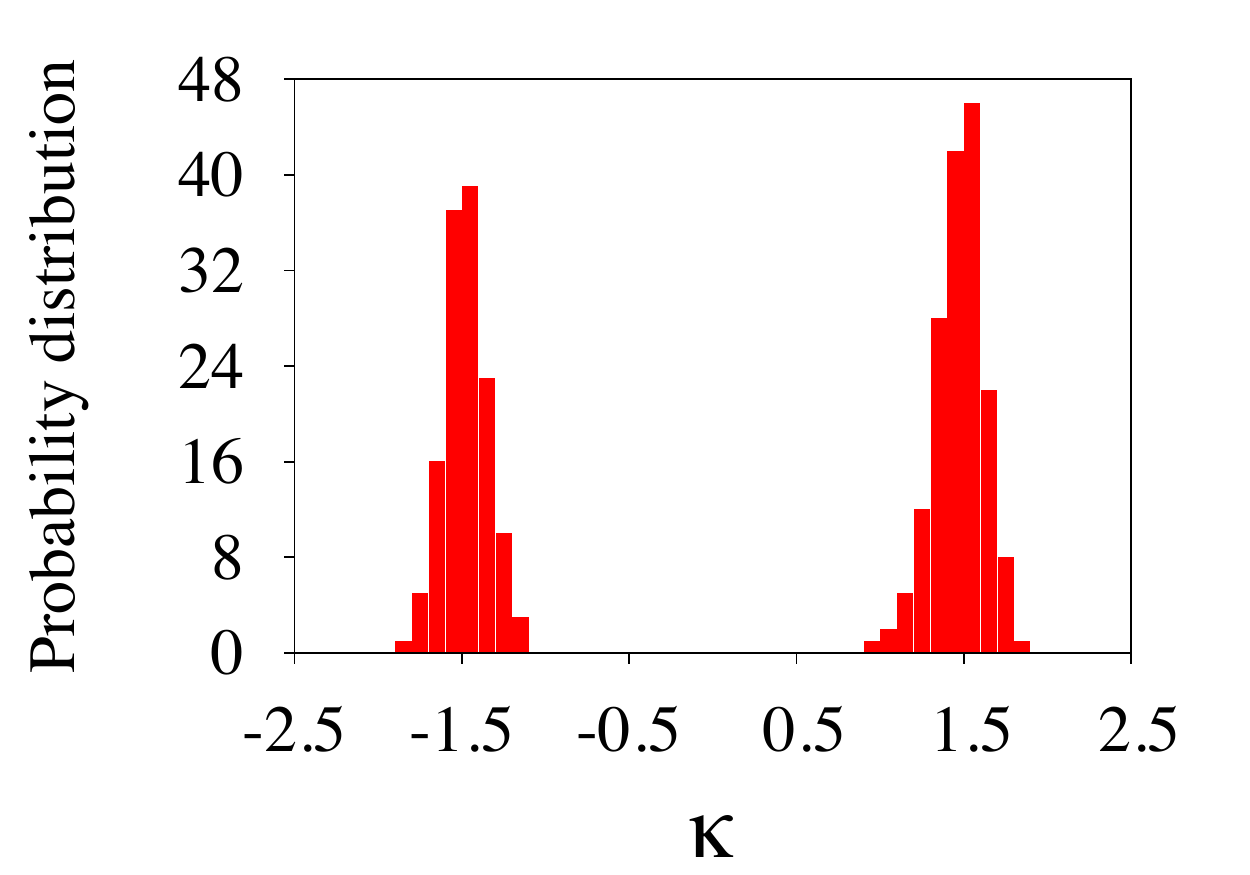}}
 \subfloat[]{\includegraphics[width=0.25\textwidth]{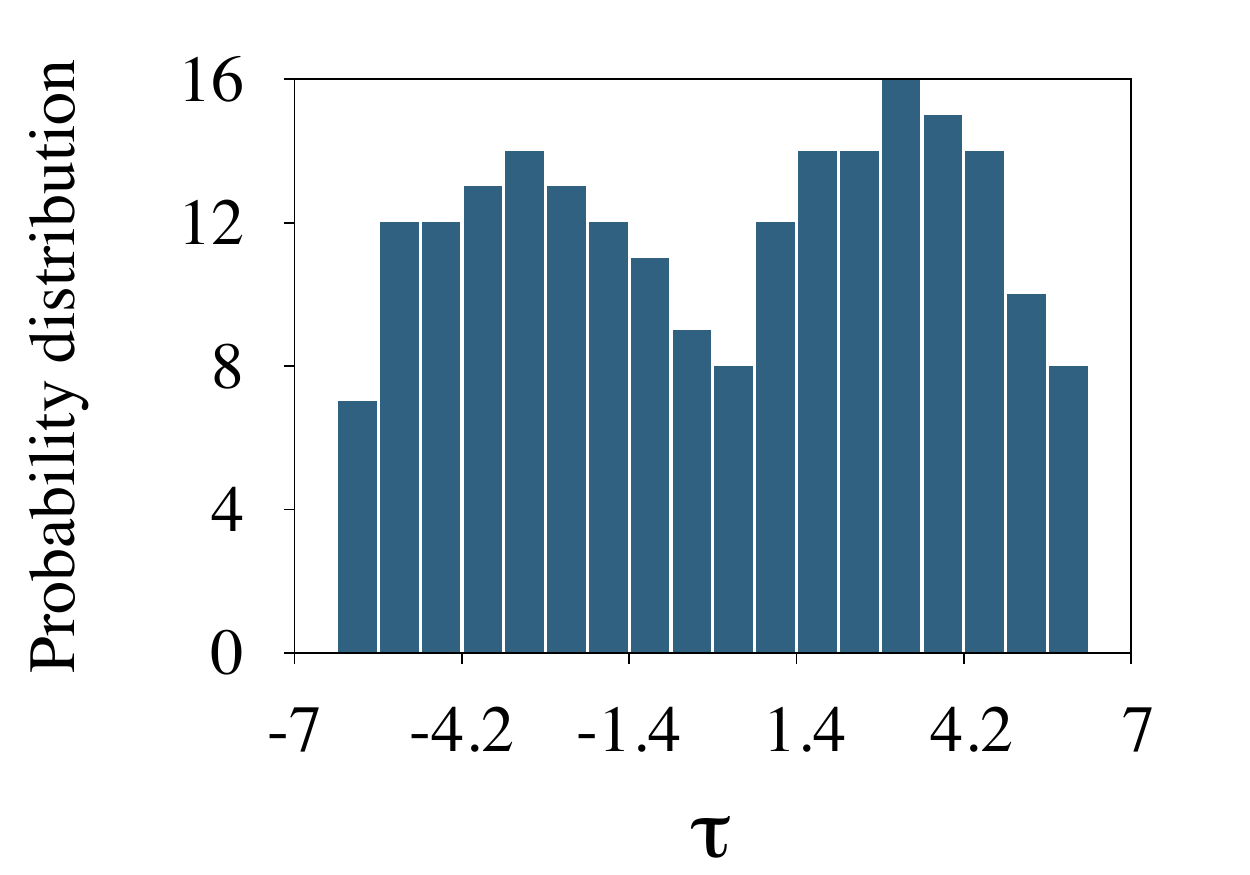}}
\vskip -0.5cm    
    \caption{Distribution of ($\kappa_i,\tau_i$) in the intermediate transition regime: an example of pseudogap state (see its description in section \ref{sec:pseudogap}. Simulation parameters are the same as for the figure 8 ($U_0=0.01$), temperature is $T=1$.}
    \label{fig-13}
  \end{figure}

\begin{figure}
\captionsetup[subfigure]{labelformat=empty}
\raggedright
 \subfloat[]{\includegraphics[width=0.25\textwidth]{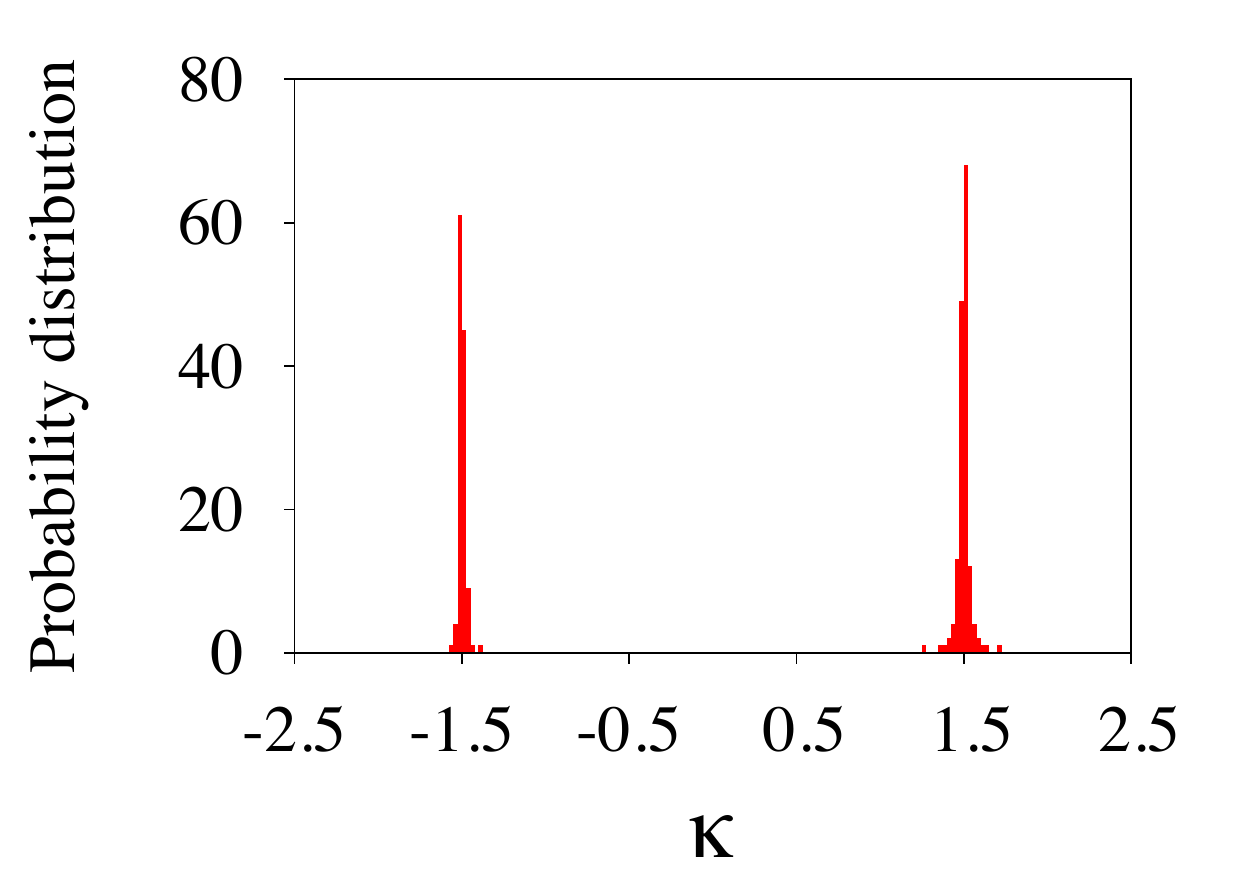}}
 \subfloat[]{\includegraphics[width=0.25\textwidth]{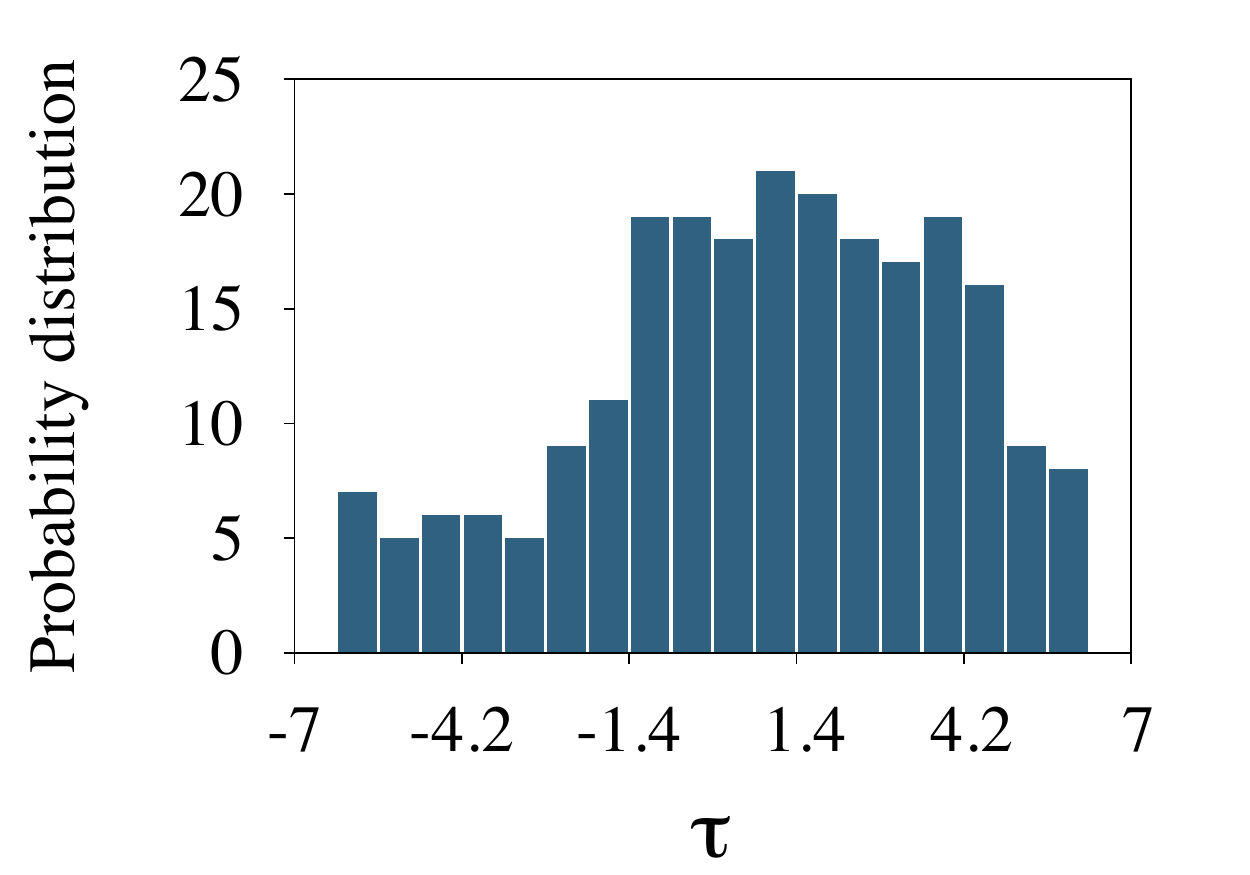}}
\vskip -0.5cm 
  \caption{Distribution of ($\kappa_i,\tau_i$) in the collapsed phase. It should be observed
  that the distribution of $\tau_i$ is asymmetric, corresponding to broken chirality. Simulation parameters are the same as for the figure 8 ($U_0=0.01$), temperature is $T=10^{-7}$.} 
    \label{fig-14}
  \end{figure}

\begin{figure}
\captionsetup[subfigure]{labelformat=empty}
\raggedright
 \subfloat[]{\includegraphics[width=0.25\textwidth]{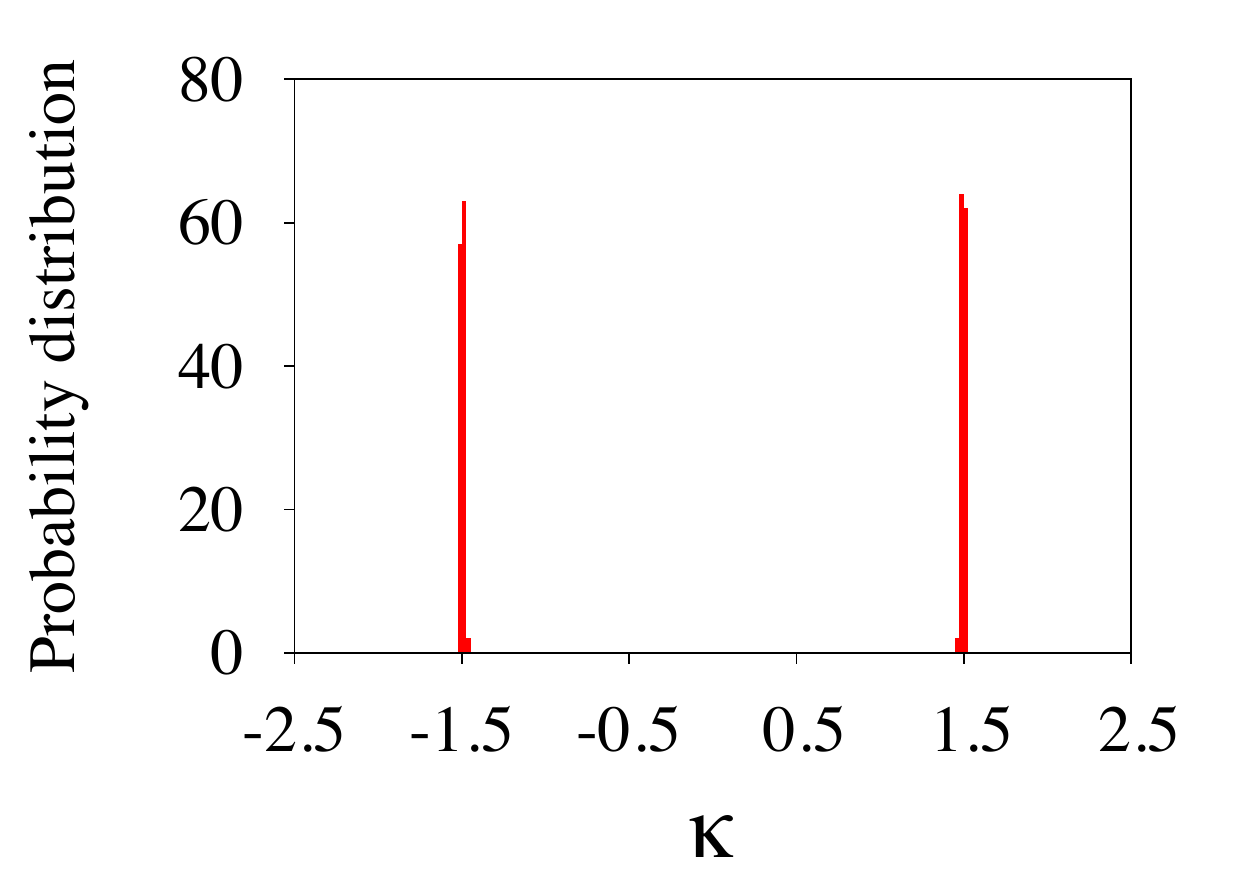}}
 \subfloat[]{\includegraphics[width=0.25\textwidth]{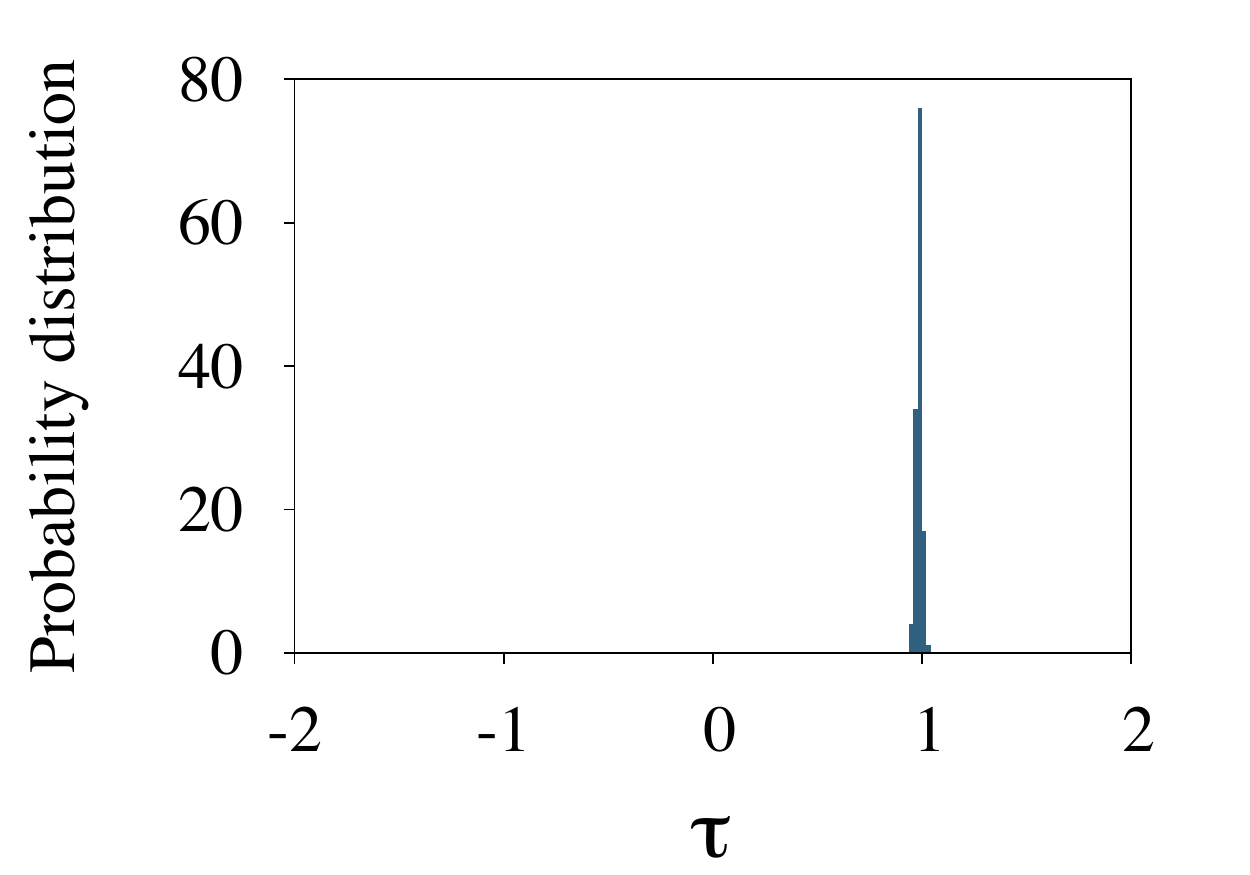}}
\vskip -0.5cm  
  \caption{Distribution of ($\kappa_i,\tau_i$) in the helical rod-like phase. Simulation parameters are the same as for the figure 12, temperature is $T=10^{-7}$.}
    \label{fig-15}
  \end{figure}

The bond and torsion angles form the complete set of  local order parameters 
to probe the  phase structure (\ref{nuval}), in the case of the present homopolymer model.  
These order parameters have the following characteristics, in the different regimes that have been 
analysed in Figures \ref{fig-7}-\ref{fig-11}; the results are summarised in Figures \ref{fig-12}-\ref{fig-15}.

In the very high temperature SAW phase, both the bond angle and the torsion angle 
are subject to large fluctuations; the simulation results  are shown in Figure \ref{fig-12}.
For the bond angles, the values are distributed in the range 
$
0 \ \leq  \ \ \langle \!\! |\kappa |\!\! \rangle \  \ \leq  \ \kappa_{max} \ \sim \ 2.2
$.
The upper limit reflects the forbidden volume constraint  (\ref{fvol}). 
As temperature increases, the values of $\kappa$ become increasingly evenly distributed over this
range so that in the $T \to \infty$ limit the distribution is fully uniform; the Figure \ref{fig-12}
shows the bond and torsion angle distribution at a generic but high temperature value. 
It is apparent  from this Figure that  both angles are {\it disordered}. 

It is concluded that the SAW phase 
is a {\it disordered phase}.
   
Next, we observe the intermediate regime takes place between
temperature values within the range
$
10^{-3} < T < 10^1 
$ 
as can be seen 
in Figures \ref{fig-7} and \ref{fig-11}. In this  intermediate region the values of the 
bond angle are found to become {\it ordered}. This is shown in  Figure \ref{fig-13}: The 
values of $\kappa_i$ are thermally fluctuating at around $|\kappa| \approx 1.5$.
The values of the torsion angle remain largely {\it disordered}.  The intermediate region is 
identified as a {\it pseudogap state},  as described in the Introduction.

Finally, there are two different low temperature phases: The collapsed phase shown in Figure
\ref{fig-7} where the attractive short-distance interaction dominates and the helical rod-like 
phase shown in Figure \ref{fig-11} where the attractive short-distance interaction becomes weak.

The distribution of bond and torsion
angles in the collapsed phase are displayed in Figure \ref{fig-14}. The bond angle is highly {\it ordered} 
around the classical value (\ref{ksolu}) but the torsion angle remains {\it disordered}. However, there is an
apparent spontaneous symmetry breaking that has taken place; the double well structure seen
in the $\tau$ distribution of Figure \ref{fig-13} has been removed,  in a way that resembles the familiar
spontaneous symmetry breaking in a  $\mathbb Z_2$ symmetric potential well.

In the helical rod-like phase, both the bond and torsion angles become peaked around the classical 
values (\ref{ksolu}). The configurations are akin straight helical rods; a little like {\it e.g.} 
collagen when biologically active. The $\kappa$ distribution
reflects the discrete $\mathbb Z_2$ gauge symmetry. But the $\mathbb Z_2$ symmetry in the $\tau$
distribution observed in Figure \ref{fig-13} is fully broken.

The transition between the collapsed phase and the helical rod-like phase entails a transitition where
the torsion angles become ordered. Due to the very low temperature values involved, the fluctuations
are strongly suppressed and a simulation becomes tedious. It is conjectured that, when the temperature
is kept in the low temperature regime, an initial
helical rod-like structure but with parameter values corresponding to the collapsed phase, is in
a {\it glassy} phase. {\it Vice versa}, an initial collapsed configuration with parameter values
in the helical rod-like phase, will eventually become subject to {\it cold denaturation}.

\subsection{Phase diagram}

The homopolymer phase  is found to depend on three relevant  scales.

\vskip 0.2cm

-- There is the extrinsic temperature scale where the values of $\kappa_i$ become ordered.
This scale can  also be controlled intrinsically, by the parameter $q$ in (\ref{hamiltonian}),  but  the details
have  not been addressed here.

-- There is temperature scale where the values of $\tau_i$ become ordered. This scale can be 
controlled intrinsically, by the parameter ratio $a/c$ in (\ref{hamiltonian}).

 -- Finally, the effects of the scale $U_0$ for the short-range attractive interactions have been
investigated.  This parameter determines an intrinsic scale that 
controls the transition temperature alternatively to the collapsed phase, or
to the helical rod-like phase.

\vskip 0.2cm

Thus, by changing the relations between the three scales  
the phase diagram of the homopolymer can be identified;  
the phase diagram is  constructed here in terms of ($T,a,U_0$). 
All the remaining parameters are fixed, and given
by the values in Table \ref{table-1}.

\begin{figure}
\centering
{\begin{minipage}[t]{.45\textwidth}
\centering
  \includegraphics[width=1\textwidth,  angle=0]{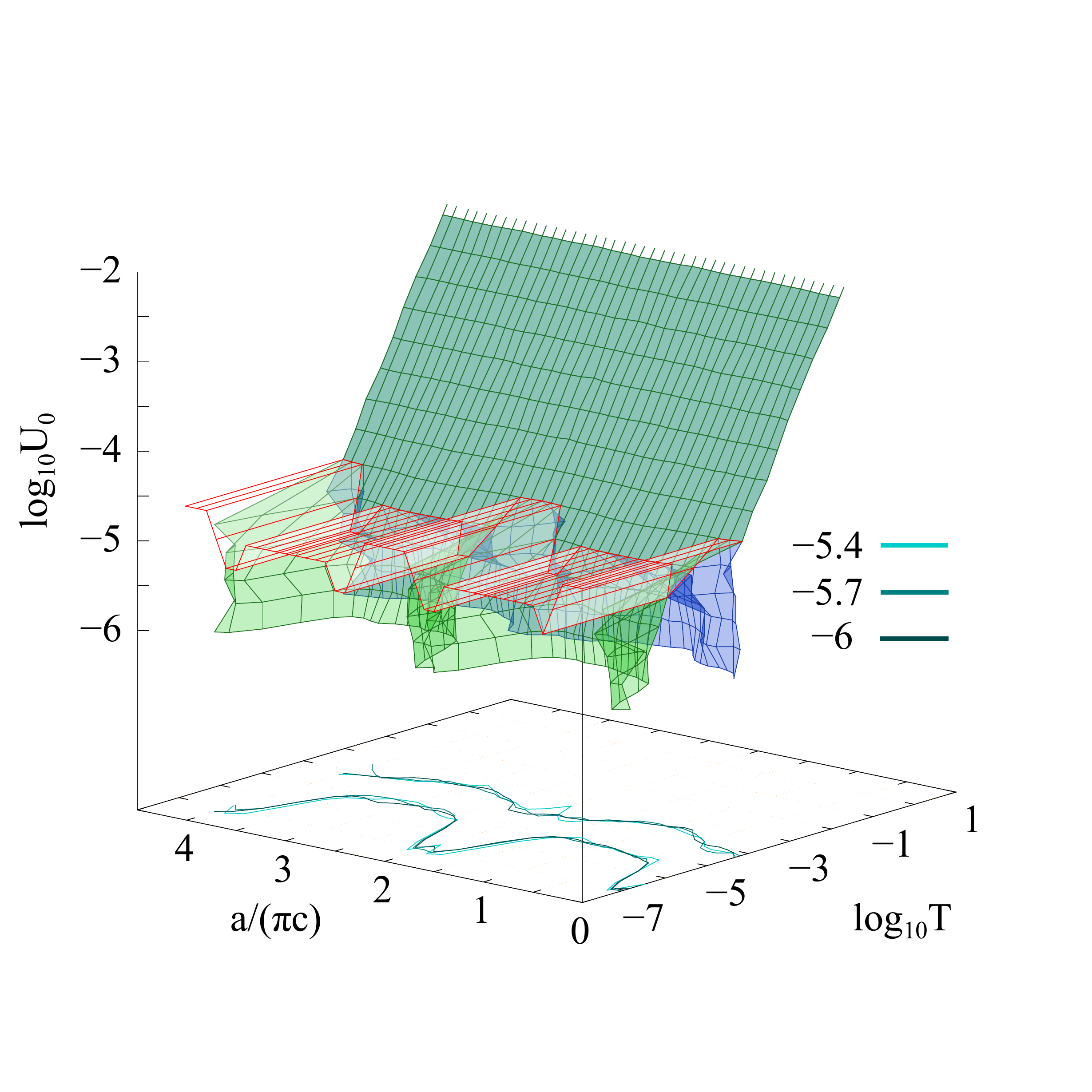}
\vskip -1cm  
  \caption{(Color online) The phase diagram on the ($U_0,T,a$) space. It was obtained for the polymer length $N=100$ using our final algorithm. All parameters in the Hamiltonian except $a$ are fixed according to the Table I.}
  \label{fig-19}
\end{minipage}}
\end{figure}

\begin{figure}
\centering
{\begin{minipage}[t]{.5\textwidth}
  \raggedright
  \includegraphics[trim = 0mm 0mm 49mm 0mm,width=.40\textwidth,  angle=0]{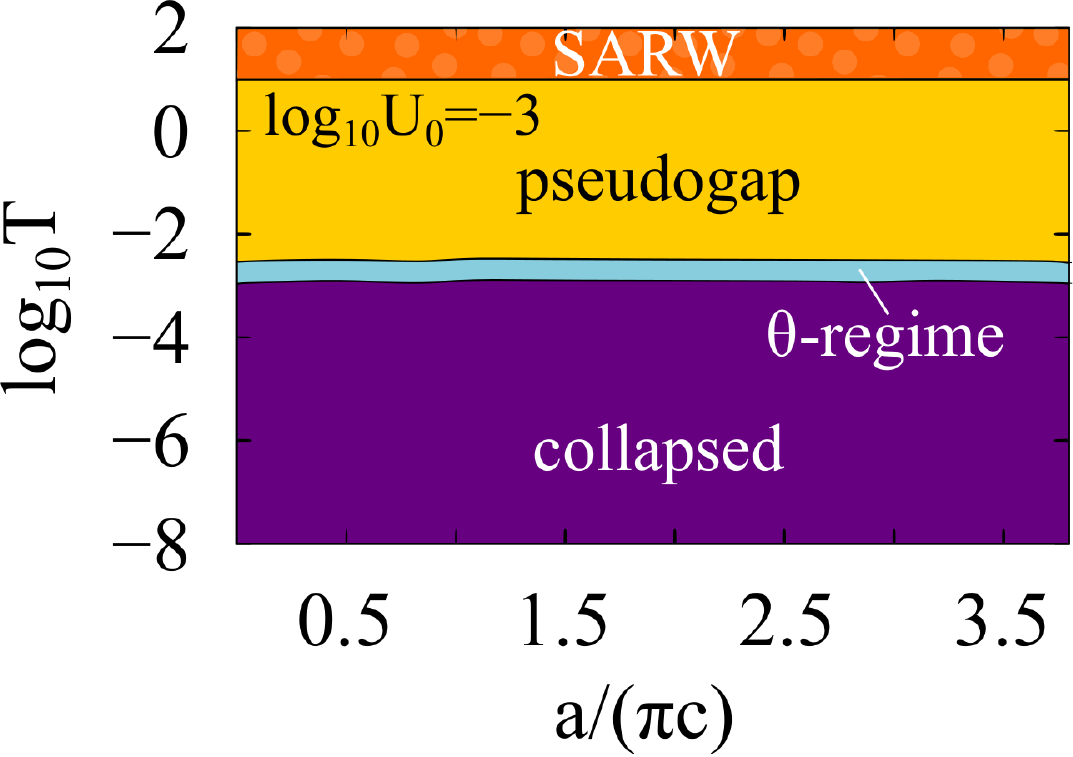}
  \caption{(Color online) A cross-section of the phase diagram in \ref{fig-14} at $U_0 = 10^{-3}$.}
  \label{fig-20}
\end{minipage}}

\centering
{\begin{minipage}[t]{.5\textwidth}
  \raggedright
  \includegraphics[trim = 0mm 0mm 49mm 0mm,width=.40\textwidth,  angle=0]{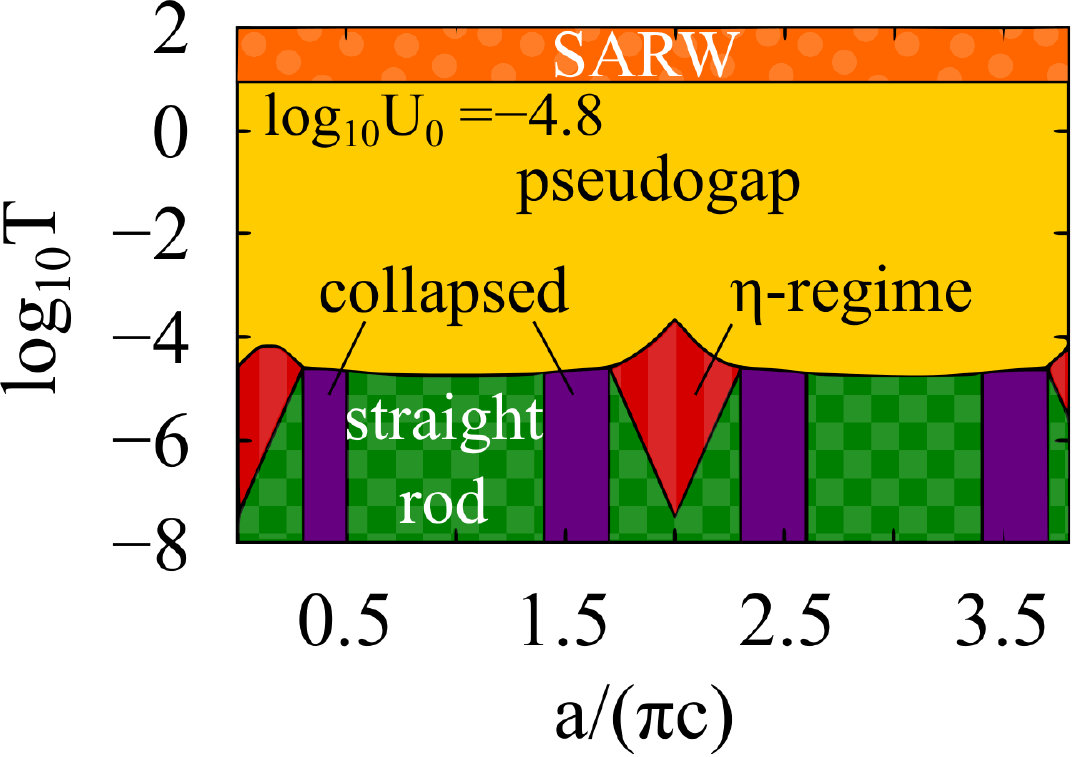}
  \caption{(Color online) A cross-section of the phase diagram in \ref{fig-14} at $U_0 = 10^{-4.8}$.}
  \label{fig-21}
\end{minipage}}

\centering
{\begin{minipage}[t]{.5\textwidth}
  \raggedright
  \includegraphics[trim = 0mm 0mm 49mm 0mm,width=.40\textwidth,  angle=0]{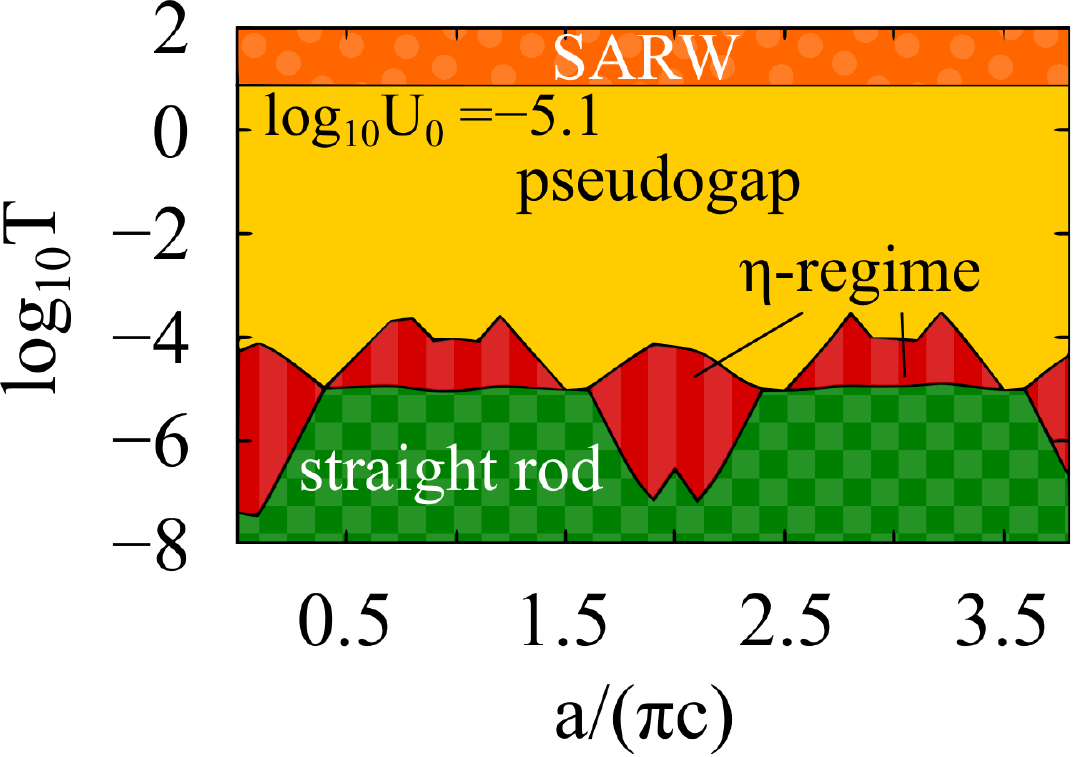}
  \caption{(Color online) A cross-section of the phase diagram in \ref{fig-14} at $U_0 = 10^{-5.1}$.}
  \label{fig-22}
\end{minipage}}

\centering
{\begin{minipage}[t]{.5\textwidth}
  \raggedright
  \includegraphics[trim = 0mm 0mm 49mm 0mm,width=.40\textwidth,  angle=0]{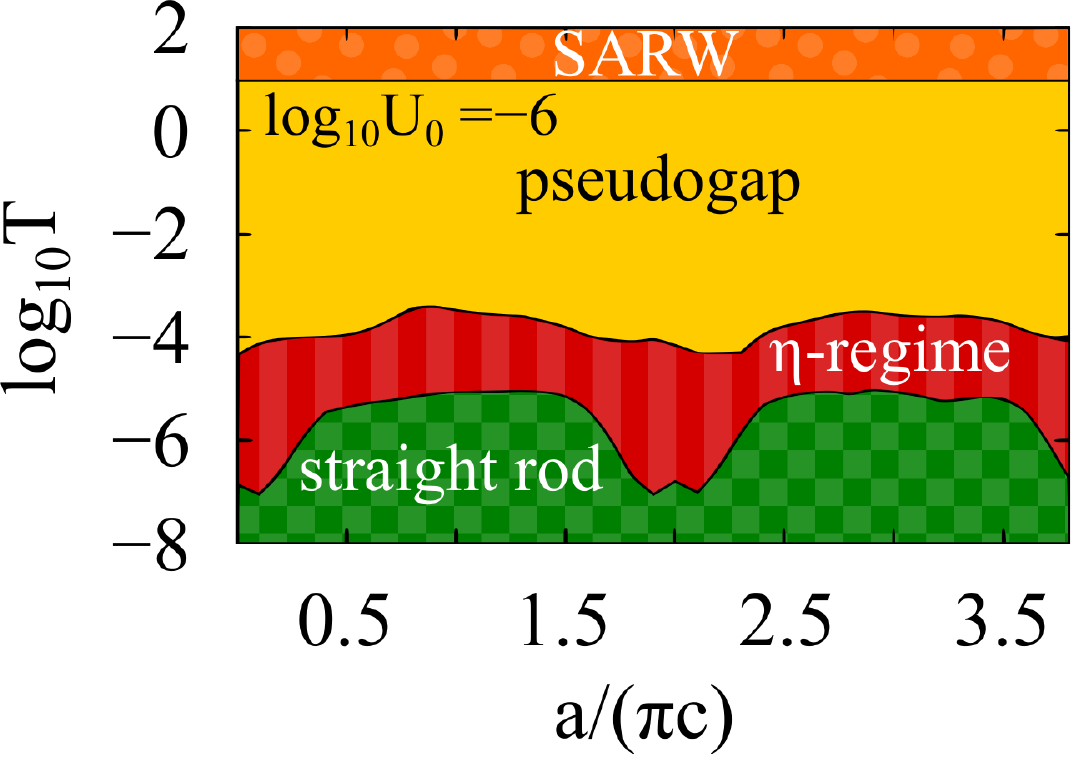}
  \caption{(Color online) A cross-section of the phase diagram in \ref{fig-14} at $U_0 = 10^{-6}$}
  \label{fig-23}
\end{minipage}}
\end{figure}

The Figure \ref{fig-19} shows the three-dimensional phase diagram in the  ($T,a,U_0$) space. The
Figures \ref{fig-20}-\ref{fig-23} show various cross-sections, taken at selected values
of $U_0$  {\it i.e.} these Figures show the phase diagram in Figure \ref{fig-19} on the ($a,T$) plane,
with different values of $U_0$.

\vskip 0.2cm

 Figure \ref{fig-20} shows the phase diagram, for a quite  large value of $U_0$ (strong coupling)  At high temperatures
there is the SAW phase. When temperature decreases there is the pseudogap state, that becomes  the collapsed
phase at low temperatures. Between the pseudogap state and the low temperature collapsed phase there is a
$\theta$-regime,  or rather a $\theta$-point as it is observed only
over a very narrow temperature range. 

-- Figure \ref{fig-21} displays the phase diagram, when the value of $U_0$  is lowered 
but still relatively large (intermediate but not weak coupling).
At high temperatures there is again the SAW phase, followed by the pseudoogap state as the temperature
decreases. At low temperatures the pseudogap state becomes converted either to the collapsed phase or
to the straight rod phase, depending on the value of the helicity parameter $a$. In addition, there is a range of 
values of $a$, when a intermediate similar to the $\theta$-regime is observed between the pseudogap state and the
straight rod phase. This is  the $\eta$-regime. It is notable, that there is a possibility of a 4-critical point
involving the pseudogap state, the $\eta$-regime, and the collapsed and straight rod phases.

-- Figure \ref{fig-22} shows the phase diagram, as the value of $U_0$  becomes further decreased
(intermediate but not strong coupling). The collapsed
state has entire disappeared, and replaced by the straight rod phase at very low temperatures. The $\eta$-regime
displays a periodic structure in the parameter $a$.  There appears to be a tri-critical point involving the
pseudogap state, the $\eta$-regime and the straight rod phase.

-- Finally, in Figure \ref{fig-23} the weak coupling $U_0$ phase diagram is displayed. The overall topology 
of the phase diagram is similar to the one in Figure \ref{fig-20}, but with the straight rod phase as the low temperature
phase instead of the collapsed phase.

\begin{figure}
{\begin{minipage}[t]{.45\textwidth}
\raggedright
  \includegraphics[width=.9\textwidth, angle=0]{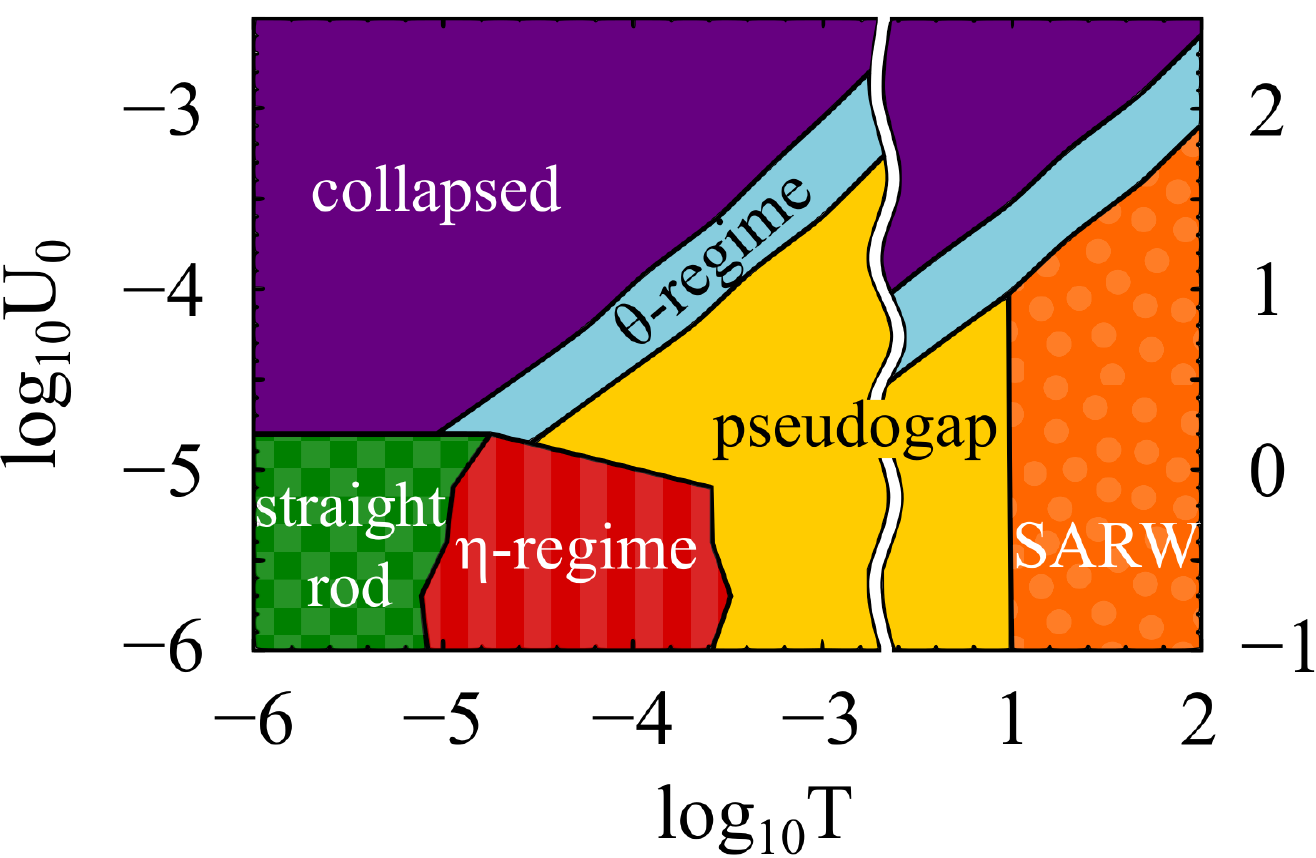}
  \caption{(Color online) The phase diagram on the ($U_0,T$) plane, with $a=10^{-4}$.}
  \label{fig-24}
\end{minipage}}
\end{figure}

\vskip 0.2cm
The Figure \ref{fig-24} shows the phase diagram on the ($U_0,T$) plane, with  helicity fixed. It is notable
that there might be a 5-criticality involving the $\eta$- and $\theta$-regimes, the pseudogap state and the
collapsed and straight rod phases. However, the detailed investigation of this region of the phase diagram is
beyond the capacity of the computer power which is presently available to us.

In summary, it is remarkable how the phase diagram (figures \ref{fig-19}-\ref{fig-23}) is periodic in the helicity
parameter $a$ (when $c$ is fixed). 
Moreover, there is a rapid transition between the helical phase
where the compactness index $\nu=1$ and the collapsed phase 
where $\nu \approx 1/3$ at low temperatures, as shown in Figure \ref{fig-21}.
There is a pseudogap state that appears as a transition regime, akin the conventional $\theta$-regime,
between the collapsed and the SAW phases. 
For the transition regime between
SAW and helical phases shown in Figure \ref{fig-22} there is a pseudogap state that 
should essentially coincide in its properties with the $\theta$-regime pseudogap state. But due to 
computational limitations the present analysis is not sufficient to confirm this.  
There is an apparent  4-critical, even 5-critical point as shown in Figures
\ref{fig-21}  and \ref{fig-24}, but  the detailed analysis of this region in the phase diagram needs to be 
performed using more extensive
simulations, which is postponed to a future project. However, it is observed that the potential presence
of a 4-critical point in a theoretical context {\it very} similar to the present one ({\it i.e.} Abelian Higgs model)
has been reported in \cite{bock-2007}

\subsection{Heteropolymer and proteins}
\label{sec:hetero}

\begin{figure}
\captionsetup[subfigure]{labelformat=empty}
\centering
\subfloat[]{\includegraphics[width=0.15\textwidth]{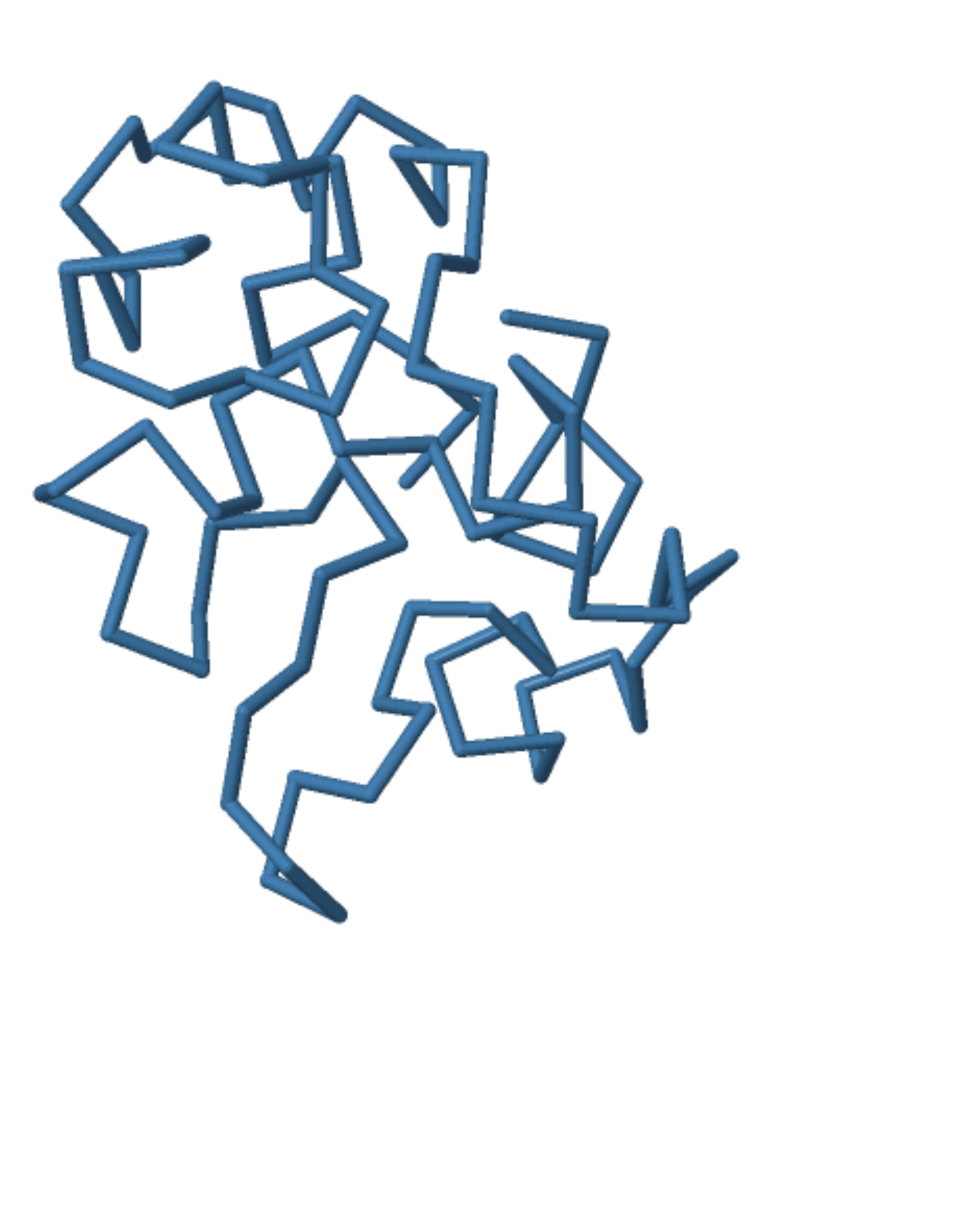}}
\subfloat[]{\includegraphics[width=0.20\textwidth]{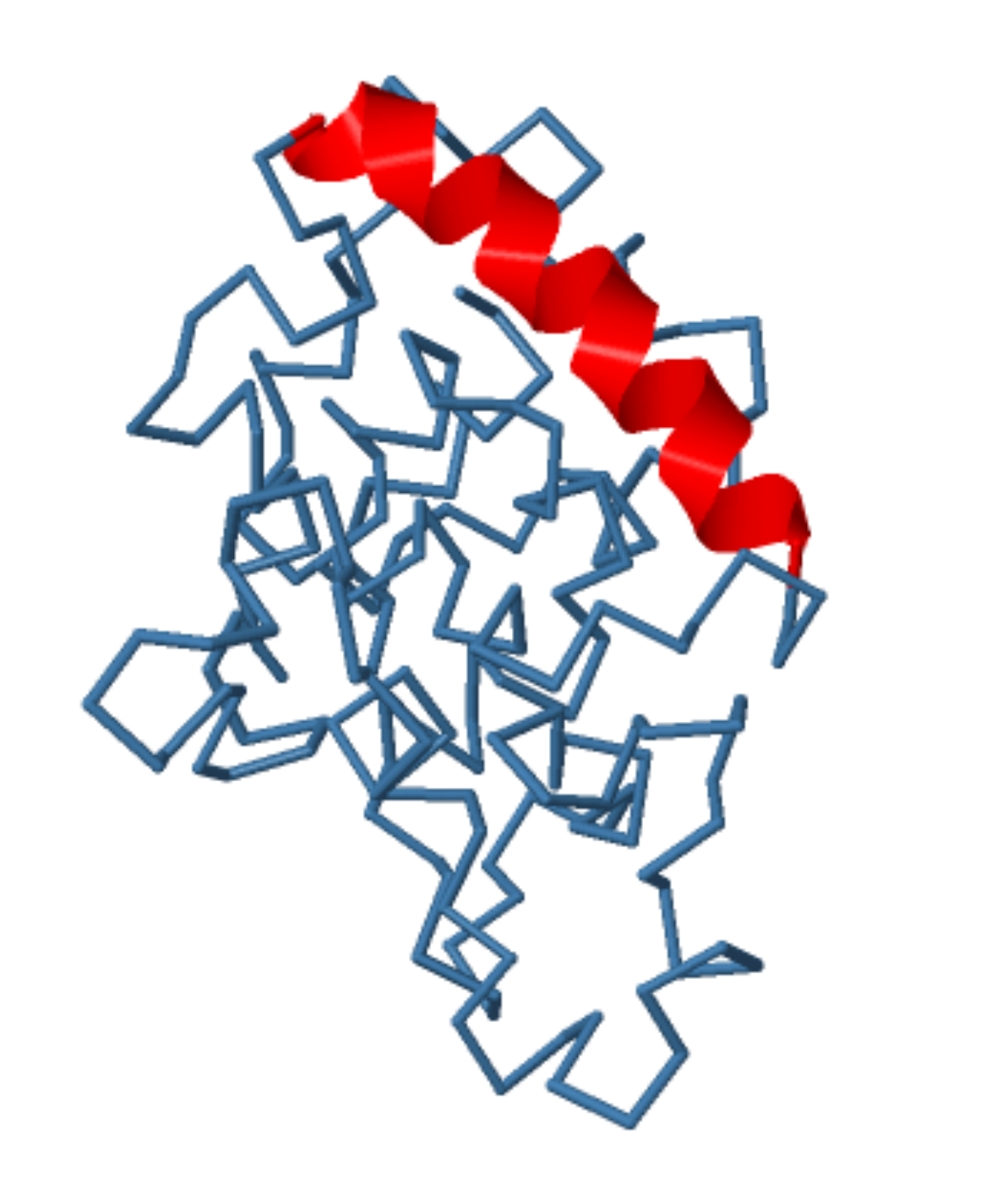}}
\vskip -0.5cm
\caption{(Color online) (left) Homopolymer does not display helices in the ground state. (right) Addition of a subchain
  that breaks homogeneity, gives rise to a helix in the subchain region. Homopolymer configuration was obtained in simulation under the parameters inside the ‘’collapsed’’ region of the phase diagram (see figure 17). In case of heteropolymer the simulation parameters are the same as for the figure 23, dimensionless temperature $T$ (see paragraph IIH) is equal to  $10^{-7.5}$.}
    \label{fig-18}
  \end{figure}

Finally,  it is inquired how the present results could be extended to heteropolymers, to draw conclusions on
the potential phase structure of proteins.  For this,
the effect of perturbations  that break the homogeneity of the  homopolymer model
have been investigated as follows: A collapse has been found to take 
place when the parameter $U_0$ that characterises the strength of self-interaction is 
larger than the parameters $a$ and $c$
that characterise the torsion angle dependent terms. Thus, a short segment is 
introduced along the chain, where $U_0$ is less 
than $a$ and $c$. Accordingly, a simulation is performed where a
heteropolymer is constructed so that for a short sub-chain of 12 monomers, the values of  the parameters $a$ and $c$
is increased from $a=c=10^{-6}$   to $a=c=10^{-2}$, while $U_0=10^{-4}$ along the entire chain.
\begin{figure}
\centering
{\begin{minipage}[t]{.5\textwidth}
  \raggedright
  \includegraphics[trim = 0mm 75mm 0mm 70mm, width=1\textwidth,  angle=0]{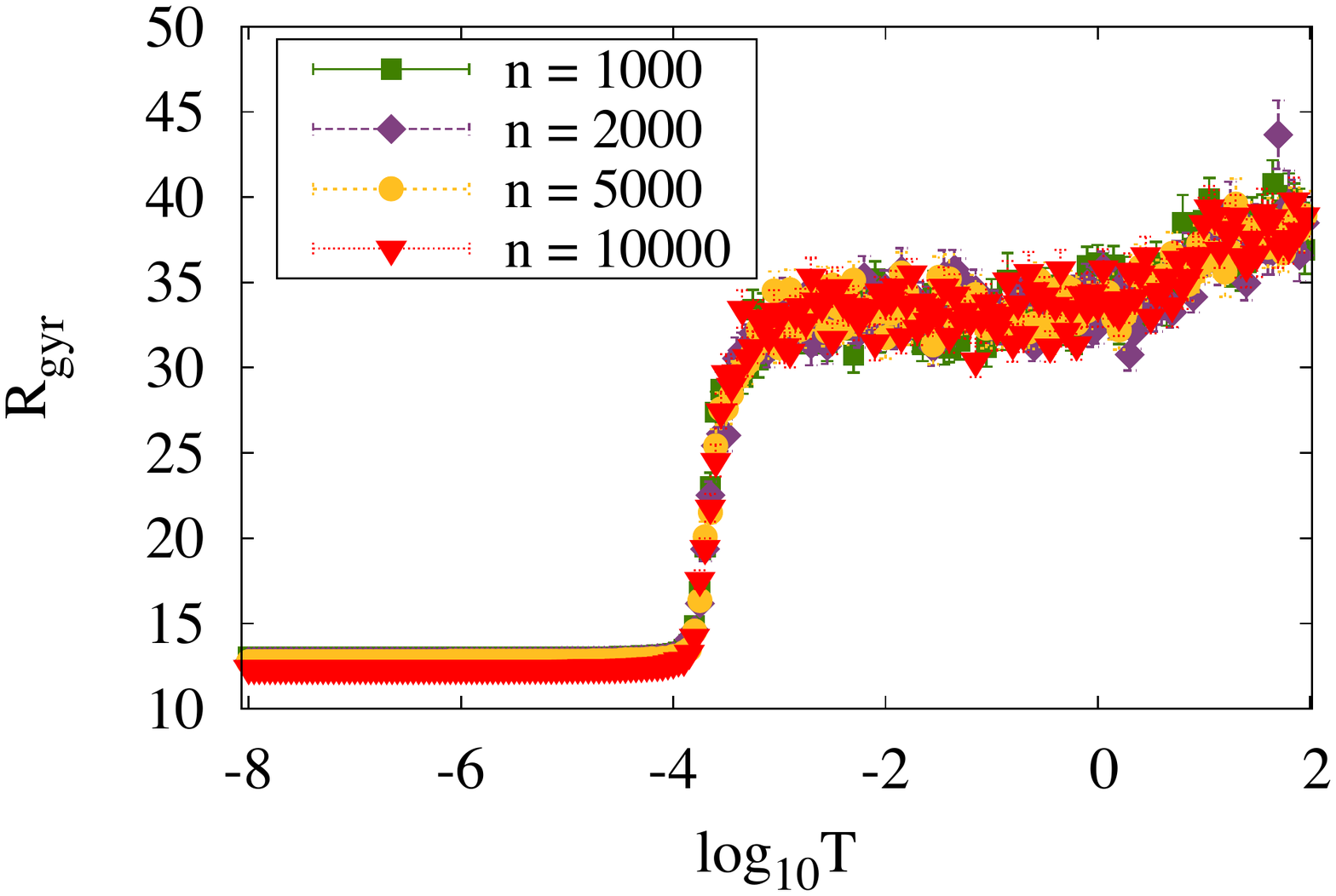}
  \caption{(Color online) Dependence of $R_{gyr}$ on temperature for a heteropolymer chain with 150 monomers 
  Thermalization length $n$ is number of updates per one step of simulated annealing. Parameters of the Hamiltonian and attraction potential are described in the paragraph D of the section III.}
  \label{fig-16}
\end{minipage}}

\noindent
{\begin{minipage}[t]{.5\textwidth}
  \raggedright
  \includegraphics[trim = 0mm 75mm 0mm 70mm, width=1\textwidth,  angle=0]{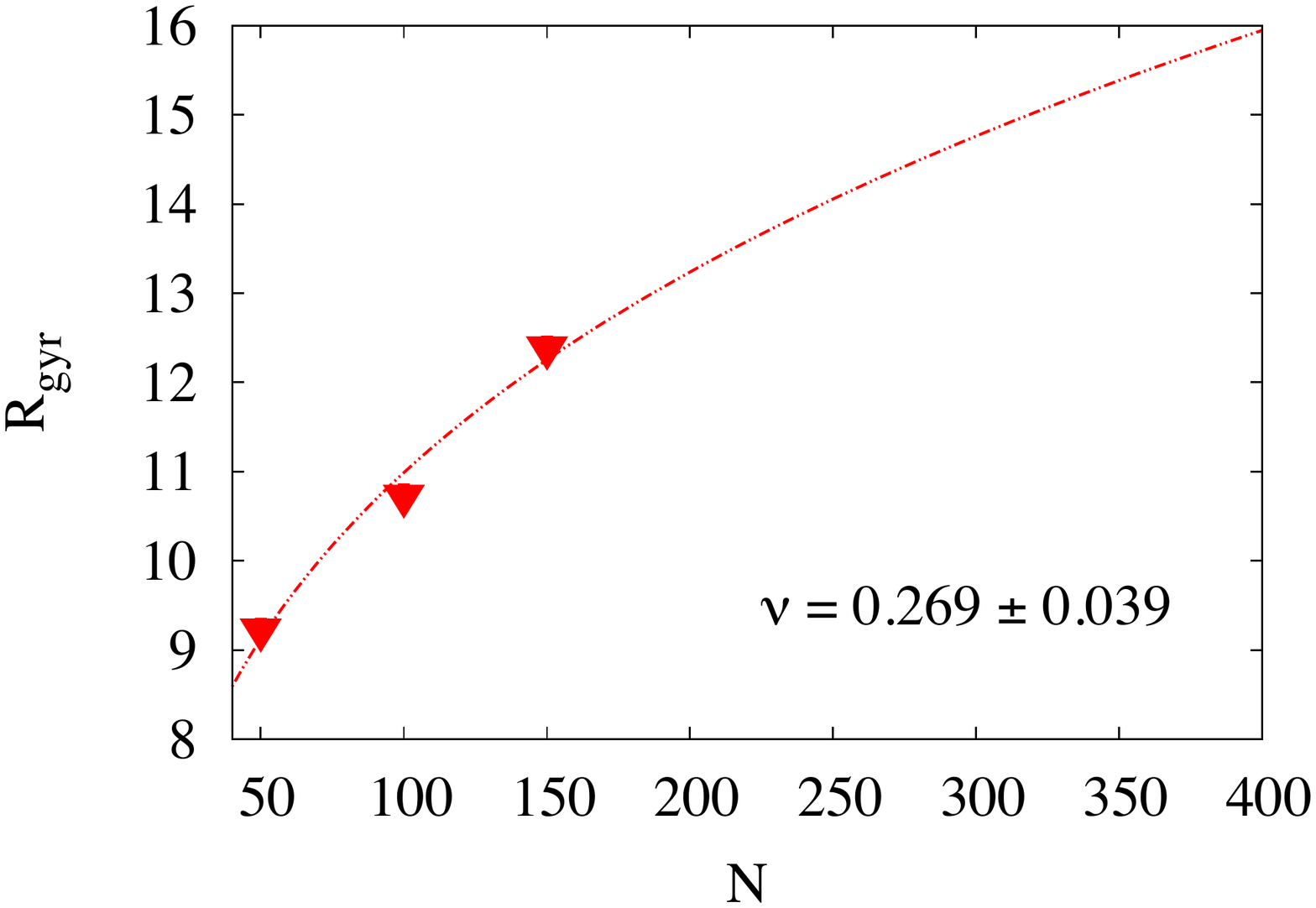}
  \caption{Dependence of $R_{gyr}$ on the length of polymer chain for low temperature $T=10^{-7.5}$. Simulation parameters are the same as for the figure 23.}
  \label{fig-17}
\end{minipage}}
\end{figure}
Results of  simulations are presented in Figures \ref{fig-16} and \ref{fig-17}, for a chain with 150 monomers. 
The dependence of $R_{gyr}$ on temperature is shown in figure \ref{fig-16};
the phase diagram is very similar to \ref{fig-7}. For example, at low temperatures it is found that the compactness index is 
close to the mean
field value $\nu = 1/3$  of the collapsed phase. However, the geometry of a configuration 
in the collapsed phase is different: As shown in Figure \ref{fig-18}
a helical structure appears only in that sub-chain where the parameter values $a$ and  $c$ have been increased.


\section{Summary}

The phase structure of chiral homopolymers under varying ambient temperature values 
have been investigated theoretically, in terms of a universal 
infrared limit energy function in combination with forbidden volume constraints (self-avoidance) and 
a short-range attractive interaction between residues. 
As such, the model should
provide a realistic description even in the case of heteropolymers that display an 
approximatively repeating monomer pattern, provided
the scale of the repeat can be considered small in comparison to the chain length. 
A biologically important example is given by
collagen, the most prevalent protein in a human body, in which case  the short range 
attractive interaction models weak hydrophobicity of the amino acids. 

It is found that the phase diagram displays a high level of complexity, in terms of the parameters that control
the helicity,  and the strength of the attractive interaction. In particular, the low energy phase is either 
like a linear one dimensional straight rod, or a  space filling collapsed configuration. At intermediate temperatures,
there is a state which can be identified as an example of the pseudogap state, and there are also 
intermediates that are more  like the conventional $\theta$-regime. It is possible that these regimes merge,
in the thermodynamical limit, at least for some range of 
parameter values. However, the possibility of the existence of 
4-critical, even 5-critical points in the phase diagram is  also proposed, but can not be confirmed with presently
available computer power.

The extension of the present approach to investigate properties of proteins and other heteropolymers remains
a challenge to future research.
 
 \vskip 0.5 cm
 
\section{Acknowledgements}
AJN  acknowledges support from Region Centre 
Rech\-erche d$^{\prime}$Initiative Academique grant, 
Vetenskapsr\aa det, Carl Trygger's Stiftelse f\"or vetenskaplig forskning, 
and  Qian Ren Grant at BIT. The work of MU was supported by the DFG grant SFB/TR-55 and by Grant
RFBR-14-02-01261-a. Numerical calculations were performed at the ITEP
computer systems ``Graphyn'' and ``Stakan'' and at the supercomputer
center of Moscow State University.  
AJN thanks A. Sieradzan and Nevena Ilieva for discussions.
We dedicate our  research to the memory of M. Polikarpov, a close colleague, mentor and teacher
without whom our collaboration would not be. Thank you, Misha.

\end{document}